\documentclass{svmult}

\usepackage{ifthen}
\newboolean{longVersion}
\setboolean{longVersion}{true}
\usepackage[pdftex]{graphicx}
\usepackage[tight,footnotesize]{subfigure}
\renewcommand*\thefigure{\thechapter.\thesection.\arabic{figure}}

\makeatletter
\renewcommand\p@subfigure{\thefigure}
\makeatother
\usepackage{caption}
\usepackage{numprint}
\usepackage{my_commands}

\numberwithin{figure}{chapter}

\begin{document}
%
%
\pagestyle{headings}  
\tableofcontents
\mainmatter              
\title{An Analysis of Anonymity in the Bitcoin System}
\titlerunning{Bitcoin Anonymity}  
%
\author{Fergal Reid \and Martin Harrigan }
\authorrunning{Reid and Harrigan} 
%
\tocauthor{Fergal Reid, Martin Harrigan}
\institute{Clique Research Cluster, \\Complex \& Adaptive Systems Laboratory,\\ University College Dublin, Ireland\\
\email{fergal.reid@gmail.com, martin.harrigan@ucd.ie}}

\maketitle              

\begin{abstract}
\\Anonymity in Bitcoin, a peer-to-peer electronic currency system, is a complicated issue. Within the system, users are identified by public-keys only. An attacker wishing to de-anonymize its users will attempt to construct the one-to-many mapping between users and public-keys and associate information external to the system with the users. Bitcoin tries to prevent this attack by storing the mapping of a user to his or her public-keys on that user's node only and by allowing each user to generate as many public-keys as required. In this chapter we consider the topological structure of two networks derived from Bitcoin's public transaction history. We show that the two networks have a non-trivial topological structure, provide complementary views of the Bitcoin system and have implications for anonymity. We combine these structures with external information and techniques such as context discovery and flow analysis to investigate an alleged theft of Bitcoins, which, at the time of the theft, had a market value of approximately half a million U.S. dollars.
\keywords{Network Analysis, Anonymity, Bitcoin}
\end{abstract}

\section{Introduction}
\label{sec:introduction}

Bitcoin is a peer-to-peer electronic currency system first described in a paper by Satoshi Nakamoto (a pseudonym) in 2008~\clubpenalty10000\cite{nakamoto-08}. It relies on digital signatures to prove ownership and a public history of transactions to prevent double-spending. The history of transactions is shared using a peer-to-peer network and is agreed upon using a proof-of-work system~\clubpenalty10000\cite{dwork-naor-92,back-02}.

The first Bitcoins were transacted in January 2009 and by June 2011 there were 6.5 million Bitcoins in circulation among an estimated 10,000 users~\clubpenalty10000\cite{the-economist-bitcoin-11}. In recent months, the currency has seen rapid growth in both media attention and market price relative to existing currencies. At its peak, a single Bitcoin traded for more than US\$30 on popular Bitcoin exchanges. At the same time, U.S. Senators and lobby groups in Germany, such as Der Bundesverband Digitale Wirtschaft (BVWD) or the Federal Association of Digital Economy, have raised concerns regarding the untraceability of Bitcoins and their potential to harm society through tax evasion, money laundering and illegal transactions. The implications of the decentralized nature of Bitcoin for authorities' ability to regulate and monitor the flow of currency is as yet unclear.

\begin{figure}[!htb]
\begin{center}
        \includegraphics[width=60mm]{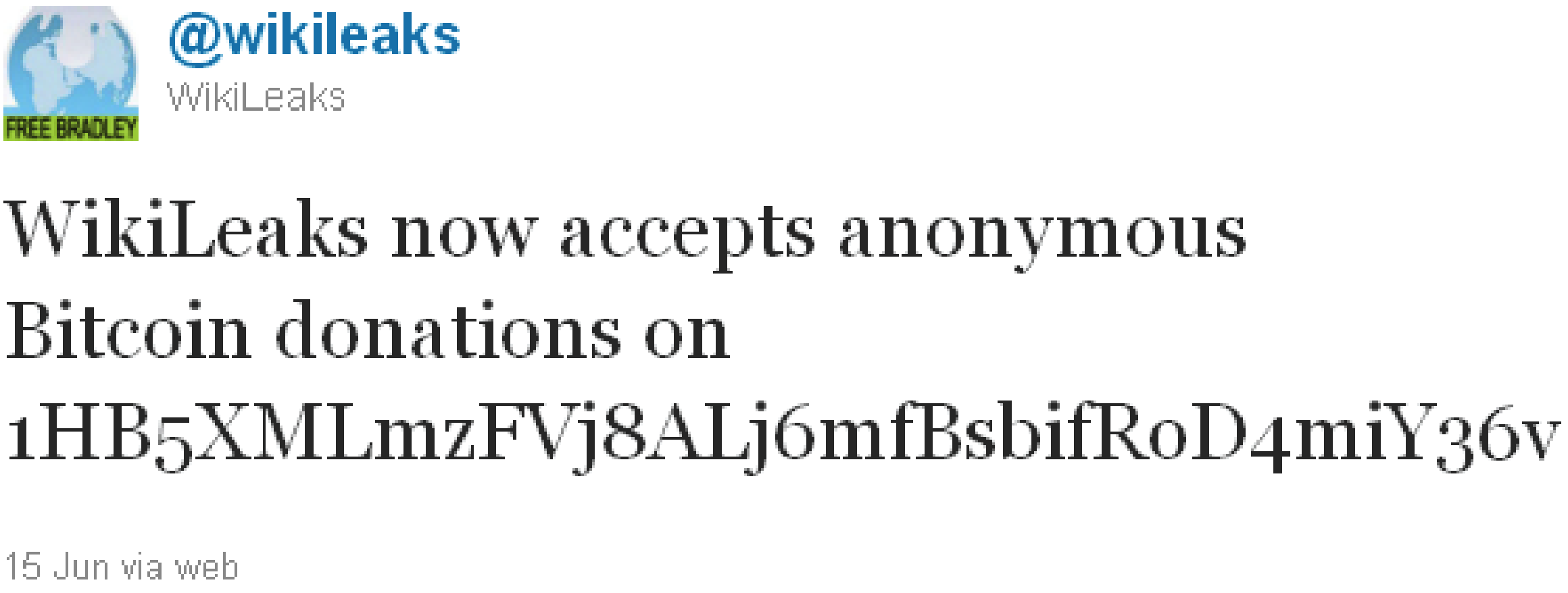}
\end{center}
\caption{Screen capture of a tweet from WikiLeaks announcing their acceptance of `anonymous Bitcoin donations'.}
\label{fig:wikileaks}
\end{figure}

Many users adopt Bitcoin for political and philosophical reasons, as much as pragmatic ones.
There is an understanding amongst Bitcoin's more technical users that anonymity is not a promenient design goal of the system; however, opinions vary widely as to how anonymous the system is, in practice.
Jeff Garzik, a member of Bitcoin's development team, is quoted as saying it would be unwise ``to attempt major illicit transactions with Bitcoin, given existing statistical analysis techniques deployed in the field by law enforcement''\footnote{http://www.theatlantic.com/technology/archive/2011/06/libertarian-dream-a-site-where-you-buy-drugs-with-digital-dollars/239776 \dd\ Retrieved 2011-11-12}; however, prior to this work, no analysis of anonymity in Bitcoin was publicly available to substantiate or refute these claims.
Furthermore, many other users of the system do not share this belief.
 For example, WikiLeaks, an international organization for anonymous whistleblowers, recently advised its Twitter followers that it now accepts \emph{anonymous} donations via Bitcoin (see Fig.~\ref{fig:wikileaks}) and states that\footnote{http://wikileaks.org/support.html \dd\ Retrieved: 2011-07-22}:
\begin{quote}
``Bitcoin is a secure and anonymous digital currency. Bitcoins cannot be easily tracked back to you, and are a [sic] safer and faster alternative to other donation methods.''
\end{quote}
They proceed to describe a more secure method of donating Bitcoins that involves the generation of a one-time public-key but the implications for those who donate using the tweeted public-key are unclear. Is it possible to associate a donation with other Bitcoin transactions performed by the same user or perhaps identify them using external information? The extent to which this anonymity holds in the face of determined analysis remains to be tested.

\ifthenelse {\boolean{longVersion}} {

This chapter is organized as follows. In Sect.~\ref{sec:related_work} we consider some existing work relating to electronic currencies and anonymity. The economic aspects of the system, interesting in their own right, are beyond the scope of this work. In Sect.~\ref{sec:the_bitcoin_system} we present an overview of the Bitcoin system; we focus on three features that are particularly relevant to our analysis. In Sect.~\ref{sec:the_transaction_and_user_networks} we construct two network structures, the transaction network and the user network using the publicly available transaction history. We study the static and dynamic properties of these networks. In Sect.~\ref{sec:anonymity_analysis} we consider the implications of these network structures for anonymity. We also combine information external to the Bitcoin system with techniques such as flow and temporal analysis to illustrate how various types of information leakage can contribute to the de-anonymization of the system's users. Finally, we conclude in Sect.~\ref{sec:conclusions}.

\subsection{A Note Regarding Motivation and Disclosure}
Our motivation for this analysis is not to de-anonymize individual users of the Bitcoin system. Rather, it is to demonstrate, using a passive analysis of a publicly available dataset, the inherent limits of anonymity when using Bitcoin. This will ensure that users do not have expectations that are not being fulfilled by the system.

In security-related research, there is considerable tension over how best to disclose vulnerabilities~\cite{cavusoglu2005emerging}. Many researchers favor full disclosure where all information regarding a vulnerability is promptly released. This enables informed users to promptly take defensive measures. Other researchers favor limited disclosure; while this provides attackers with a window in which to exploit uninformed users, a mitigation strategy can be prepared and implemented before public announcement, thus limiting damage, \eg through a software update. Our analysis illustrates some potential risks and pitfalls with regard to anonymity in the Bitcoin system. However, there is no central authority which can fundamentally change the system's behavior. Furthermore, it is not possible to mitigate analysis of the existing transaction history.

There are also two noteworthy features of the dataset when compared with, say, contentious social network datasets, \eg the Facebook profiles of Harvard University students~\clubpenalty10000\cite{lewis-et-al-08}. Firstly, the delineation between what is considered public and private is clear: the entire history of Bitcoin transactions is publicly available. Secondly, the Bitcoin system does not have a usage policy. After joining Bitcoin's peer-to-peer network, a client can freely request the entire history of Bitcoin transactions; there is no crawling or scraping required.

Thus, we believe the best strategy to minimise the threat to user anonymity is to be descriptive about the risks of the Bitcoin system. We do not identify individual users -- apart from those in the case study -- but we note that it is not difficult for other groups to replicate our work. Indeed, given the passive nature of the analysis, other parties may already be conducting similar analyses.

} {

In this chapter we review existing work relating to electronic currencies and anonymity in Sect.~\ref{sec:related_work}, we present an overview of the Bitcoin system in Sect.~\ref{sec:the_bitcoin_system}, we detail the construction of two network structures in Sect.~\ref{sec:the_transaction_and_user_networks} and, in Sect.~\ref{sec:anonymity_analysis}, we consider the implications of these network structures, combined with external information for anonymity in the Bitcoin system.

}

\section{ Related Work }
\label{sec:related_work}

The related work for this chapter can be categorized into two fields:
electronic currencies and anonymity.

\subsection{ Electronic Currencies }
\label{sec:related_work_electronic_currencies}

Electronic currencies can be technically classified according to their
mechanisms for establishing ownership, protecting against double-spending,
ensuring anonymity and/or privacy, and generating and issuing new currency. Bitcoin is
particularly noteworthy for the last of these mechanisms. The proof-of-work
system~\clubpenalty10000\cite{dwork-naor-92,back-02} that establishes
consensus regarding the history of transactions also doubles as a minting
mechanism. The scheme was first outlined in the B-Money
Proposal~\clubpenalty10000\cite{dai-98}. We briefly consider some
alternative mechanisms. Ripple~\clubpenalty10000\cite{fugger-04} is an
electronic currency where every user can issue currency. However, the
currency is only accepted by peers who trust the issuer. Transactions
between arbitrary pairs of users require chains of trusted intermediaries
between the users. Saito~\clubpenalty10000\cite{saito-06} formalized and
implemented a similar system, i-WAT, in which the the chain of
intermediaries can be established without their immediate presence using
digital signatures. KARMA~\clubpenalty10000\cite{vishnumurthy-et-al-03} is
an electronic currency where the central authority is distributed over a
set of users that are involved in all transactions.
PPay~\clubpenalty10000\cite{yang-garcia-molin-03} is a micropayment scheme
for peer-to-peer systems where the issuer of the currency is responsible
for keeping track of it. However, both KARMA and PPay may incur a large
overhead when the rate of transactions is high. Mondex is a smart-card
electronic currency~\clubpenalty10000\cite{stalder-02}. It preserves a
central bank's role in the generation and issuance of electronic currency.
Mondex was an electronic replacement for cash in the physical world whereas
Bitcoin is an electronic analog of cash in the online world.

\ifthenelse {\boolean{longVersion}} {

The authors are not aware of any studies of the network structure of
electronic currencies. However, there are such studies of physical
currencies. The community currency Tomamae-cho was introduced into the
Hokkaido Prefecture in Japan for a three-month period during 2004--05 in a
bid to revitalize local economy. The Tomamae-cho system involved
gift-certificates that were re-usable and legally redeemable into yen.
There was an entry space on the reverse of each certificate for recipients
to record transaction dates, their names and addresses, and the purposes of
use, up to a maximum of five recipients.
Kichiji and Nishibe~\clubpenalty10000\cite{kichiji-nishibe-08} used the collected certificates
to derive a network structure that represented the flow of currency during
the period. They showed that the cumulative degree distribution of the
network obeyed a power-law distribution, the network had small-world
properties (the average clustering coefficient was high whereas the average
path length was low), the directionality and the value of transactions were
significant features, and the double-triangle
system~\clubpenalty10000\cite{nishibe-04} was effective. There also exist
studies of the physical movement of currency: `Where's
George?'~\clubpenalty10000\cite{where's-george-xx} is a crowd-sourced
method for tracking U.S. dollar bills where users record the serial numbers
of bills in their possession, along with their current location. If a bill
is recorded sufficiently often, its geographical movement can be tracked
over time. Brockmann et al.~\clubpenalty10000\cite{brockmann-et-al-06} used this dataset as
a proxy for studying multi-scale human mobility and as a tool for computing
geographic borders inherent to human mobility.

Grinberg~~\clubpenalty10000\cite{grinberg-11} considers some of the legal issues that may be relevant to Bitcoin in the United States. For example, does Bitcoin violate the Stamp Payments Act of 1862? The currency can be used as a token for ``a less sum than \$1, intended to circulate as money or to be received or used in lieu of lawful money of the United States''. However, the authors of the act could not have conceived of digital currencies at the time of its writing and therefore Bitcoin may not fall under its scope. Grinberg believes that Bitcoin is unlikely to be a security or more specifically an ``investment contract'' and therefore does not fall under the Securities Act of 1933. He also believes that the Bank Secrecy Act of 1970 and the Money Laundering Control Act of 1986 pose the greatest risk for Bitcoin developers, exchanges, wallet providers, mining pool operators and businesses that accept Bitcoins. These acts require certain kinds of financial businesses, even if they are located abroad, to register with a bureau of the United States Department of the Treasury known as the Financial Crimes Enforcement Network (or FinCEN). The legality of Bitcoin is outside the scope of our work but is interesting nonetheless.

}{

The authors are not aware of any studies of the network structure of
electronic currencies. However, there are such studies of physical
currencies, for example, Kichiji and Nishibe~\clubpenalty10000\cite{kichiji-nishibe-08} studied the flow of the community currency Tomamae-cho for a three-month period during 2004--05 and Brockmann et al.~\clubpenalty10000\cite{brockmann-et-al-06} studied the geographical movement of U.S. dollar bills.

}

\subsection{ Anonymity }
\label{sec:related_work_anonymity}

Previous work has shown the difficulty in maintaining anonymity in the context of networked data and online services which expose partial user information. Narayanan and Shmatikov~\cite{narayanan2009anonymizing} and Backstrom et al.~\cite{backstrom2007wherefore} consider privacy attacks which identify users using the structure of networks and show the difficulty in guaranteeing anonymity in the presence of network data. Crandall et al.~\cite{crandall2010inferring} infer social ties between users where none are explicitly stated by looking at patterns of `co-incidences' or common off-network co-occurrences. Gross and Acquisiti \cite{gross2005information} discuss privacy of early users in the Facebook social network, and how information from multiple sources could be combined to identify pseudonymous network users. Narayanan and Shmatikov~\cite{narayanan2008robust} de-anonymized the Netflix Prize dataset using information from IMDB\footnote{http://www.imdb.com} which had similar user content, showing that statistical matching between different but related datasets can be used to attack anonymity. Puzis et al.~\cite{puzis2009collaborative} simulated the monitoring of a communications network using strategically-located monitoring nodes and showed that, using real-world network topologies, a relatively small number of nodes can collaborate to pose a significant threat to anonymity. Korolova et al. \cite{korolova2008link} study strategies for efficiently compromising network nodes, to maximise link information observed. Altshuler et al. \cite{altshuler2011stealing} discuss the increasing dangers of attacks targeting similar types of information, and provide measures of the difficulty of such attacks, on particular networks. All of this work points to the difficulty in maintaining anonymity where network data on user behaviour is available and illustrates how seemingly minor information leakages can be aggregated to pose significant risks. 
The security researcher Dan Kaminsky independently performed an investigation of some aspects of anonymity in the Bitcoin system, which he presented at a security conference~\cite{kaminsky} shortly after an initial draft of this work was made public.
His work investigates the `linking problem' we analyze in Sect.~\ref{sec:the_user_network}.
In addition to the analysis we conducted, his work investigates the Bitcoin system from an angle we did not consider in our investigation -- the TCP/IP operation of the underlying peer-to-peer network.
Kaminsky's TCP/IP layer findings strengthen the core claims of this work that Bitcoin does not anonymise user activity. 
We provide a summary of his findings in Sect.~\ref{sec:tcp_ip_layer_information}.

\section{ The Bitcoin System }
\label{sec:the_bitcoin_system}

The following is a simplified description of the Bitcoin system; see
Nakamoto \cite{nakamoto-08} for a more thorough treatment. Bitcoin
is an electronic currency with no central authority or issuer. There is no
central bank or fractional reserve system controlling the supply of
Bitcoins. Instead, they are generated at a predictable rate such that 
the eventual total number will be 21 million. There is no requirement for a
trusted third-party when making transactions. Suppose Alice wishes to
`send'  a number of Bitcoins to Bob. Alice uses a Bitcoin client to join
the Bitcoin peer-to-peer network and makes a public transaction or
declaration stating that one or more identities that she controls (which
can be verified using public-key cryptography), and which previously had a
number of Bitcoins assigned to them, wish to re-assign those Bitcoins to
one or more other identities, at least one of which is controlled by Bob.
The participants of the peer-to-peer network form a collective consensus
regarding the validity of this transaction by appending it to the public
history of previously agreed-upon transactions (the
\emph{block-chain}). This process involves the
repeated computation of a cryptographic hash function so that the digest of
the transaction, along with other pending transactions, and an arbitrary
nonce, has a specific form. This process is designed to require considerable
computational effort, from which the security of the Bitcoin mechanism is 
derived.  To encourage users to pay this computational cost, the process is 
incentivized using newly generated Bitcoins and/or transaction fees, and so this 
whole process is known as \emph{mining}.

In this chapter, there are three features of the Bitcoin system that are of
particular interest. Firstly, the entire history of Bitcoin transactions is
publicly available. This is necessary in order to validate transactions and
prevent double-spending in the absence of a central authority. The only way
to confirm the absence of a previous transaction is to be aware of all
previous transactions. The second feature of interest is that a transaction 
can have multiple inputs and
multiple outputs. An input to a transaction is either the output of a
previous transaction or a sum of newly generated Bitcoins and transaction
fees. A transaction frequently has either a single input from a previous
larger transaction or multiple inputs from previous smaller transactions.
Also, a transaction frequently has two outputs: one sending payment and one
returning change. Thirdly, the payer and payee(s) of a transaction are
identified through public-keys from public-private key-pairs. However, a
user can have multiple public-keys. In fact, it is considered good practice
for a payee to generate a new public-private key-pair for every
transaction. Furthermore, a user can take the following steps to better
protect their identity: they can avoid revealing any identifying
information in connection with their public-keys; they can repeatedly send
varying fractions of their Bitcoins to themselves using multiple (newly
generated) public-keys; and/or they can use a trusted third-party mixer or
laundry. However, these practices are not universally applied.

The three features above, namely the public availability of Bitcoin
transactions, the input-output relationship between transactions and the
re-use and co-use of public-keys, provide a basis for two distinct network
structures: the \emph{transaction network} and the \emph{user network}. The
transaction network represents the flow of Bitcoins between
\emph{transactions} over time. Each vertex represents a transaction and
each directed edge between a source and a target represents an output of
the transaction corresponding to the source that is an input to the
transaction corresponding to the target. Each directed edge also includes a
value in Bitcoins and a timestamp. The user network represents the flow of
Bitcoins between \emph{users} over time. Each vertex represents a user and
each directed edge between a source and a target represents an input-output
pair of a single transaction where the input's public-key belongs to the
user corresponding to the source and the output's public-key belongs to the
user corresponding to the target. Each directed edge also includes a value
in Bitcoins and a timestamp.

We gathered the entire history of Bitcoin transactions from the first
transaction on the 3\superscript{rd} January 2009 up to and including the
last transaction that occurred on the 12\superscript{th} July 2011. We
gathered the dataset using the Bitcoin
client\footnote{http://www.bitcoin.org} and a modified version of
Gavin Andresen's \texttt{bitcointools} project.\footnote{http://github.com/gavinandresen/bitcointools}
The dataset comprises $\numprint{1019486}$ transactions between
$\numprint{1253054}$ unique public-keys. We describe the construction of
the corresponding transaction and user networks and their analyses in the
following sections. We will show that the two networks are complex, have a non-trivial
topological structure, provide complementary views of the Bitcoin system
and have implications for the anonymity of users.

\section{ The Transaction and User Networks }
\label{sec:the_transaction_and_user_networks}

\subsection{ The Transaction Network }
\label{sec:the_transaction_network}

The transaction network $\mathcal{T}$ represents the flow of Bitcoins
between \emph{transactions} over time. Each vertex represents a transaction
and each directed edge between a source and a target represents an output
of the transaction corresponding to the source that is an input to the
transaction corresponding to the target. Each directed edge also includes a
value in Bitcoins and a timestamp. It is a straight-forward task to
construct $\mathcal{T}$ from our dataset.

\begin{figure}\centerline{\includegraphics[width=7.5cm]{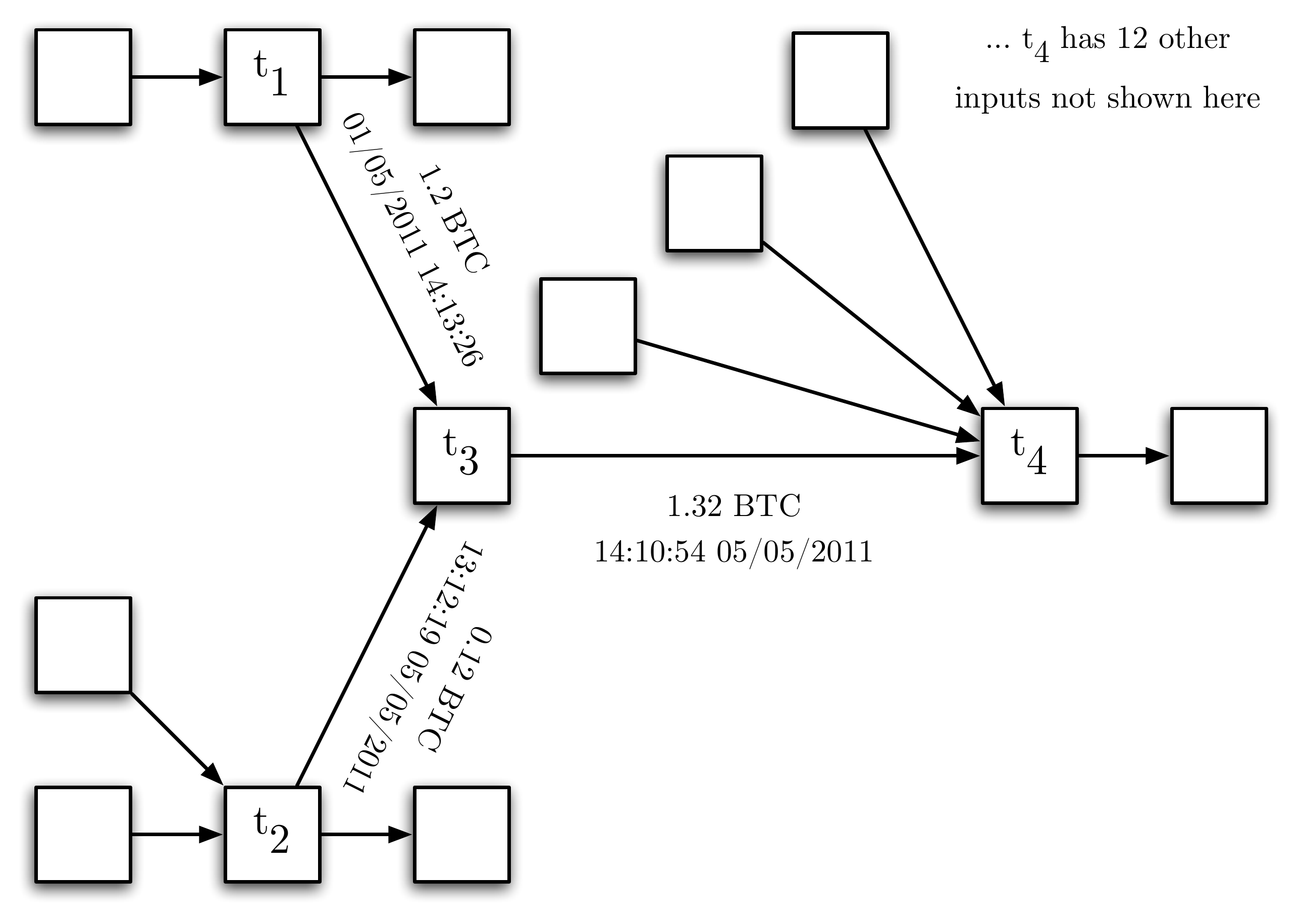}}\caption{An
example sub-network from the transaction network. Each rectangular vertex represents a
transaction and each directed edge represents a flow of Bitcoins from an
output of one transaction to an input of
another.}\label{fig:transaction_network_example}

\vspace{-10px}
\end{figure}

Figure~\ref{fig:transaction_network_example} shows an example sub-network of $\mathcal{T}$.
$t_1$ is a transaction with one input
and two outputs.\footnote{The transactions and public-keys used in our examples
exist in our dataset. The unique identifier for the transaction $t_1$ is
09441d3c52fa0018365fcd2949925182f6307322138773d52c201f5cc2bb5976. 
You can query the details of a transaction or public-key by examining
Bitcoin's block-chain using, say, the Bitcoin Block Explorer
({http://www.blockexplorer.com}).} It was added to the block-chain on the 1\superscript{st}
May 2011. One of its outputs assigned 1.2 BTC (Bitcoins) to
a user identified by the public-key
$pk_1$.\footnote{13eBhR3oHFD5wkE4oGtrLdbdi2PvK3ijMC} The public-keys are not shown in Fig.~\ref{fig:transaction_network_example}. Similarly,
$t_2$ is a transaction with two inputs and two outputs.\footnote{0c4d41d0f5d2aff14d449daa550c7d9b0eaaf35d81ee5e6e77f8948b14d62378} It was accepted on the
5\superscript{th} May 2011. One of its outputs sent 0.12
BTC to a user identified by a different public-key,
$pk_2$.\footnote{19smBSUoRGmbH13vif1Nu17S63Tnmg7h9n}
$t_3$ is a transaction with two inputs and one output.\footnote{0c034fb964257ecbf4eb953e2362e165dea9c1d008032bc9ece5cebbc7cd4697} It was accepted on the
5\superscript{th} May 2011. Both of its inputs are connected to
the two aforementioned outputs of $t_1$ and $t_2$. The only output of $t_3$ was
redeemed by
$t_4$.\footnote{f16ece066f6e4cf92d9a72eb1359d8401602a23990990cb84498cdbb93026402}

$\mathcal{T}$ has $\numprint{974520}$ vertices and $\numprint{1558854}$
directed edges. The number of vertices is less than the total number of
transactions in the dataset because we omit transactions that are not
connected to at least one other transaction. These correspond to newly
generated Bitcoins and transactions fees that are not yet redeemed. The
network has neither multi-edges (multiple edges between the same pair of
vertices in the same direction) nor loops. It is a directed acyclic graph
(DAG) since the output of a transaction can never be an input (either
directly or indirectly) to the same transaction.

\ifthenelse {\boolean{longVersion}}{

\begin{figure}[!h]
\centerline{
\subfigure[]{
\includegraphics[width=6.0cm]{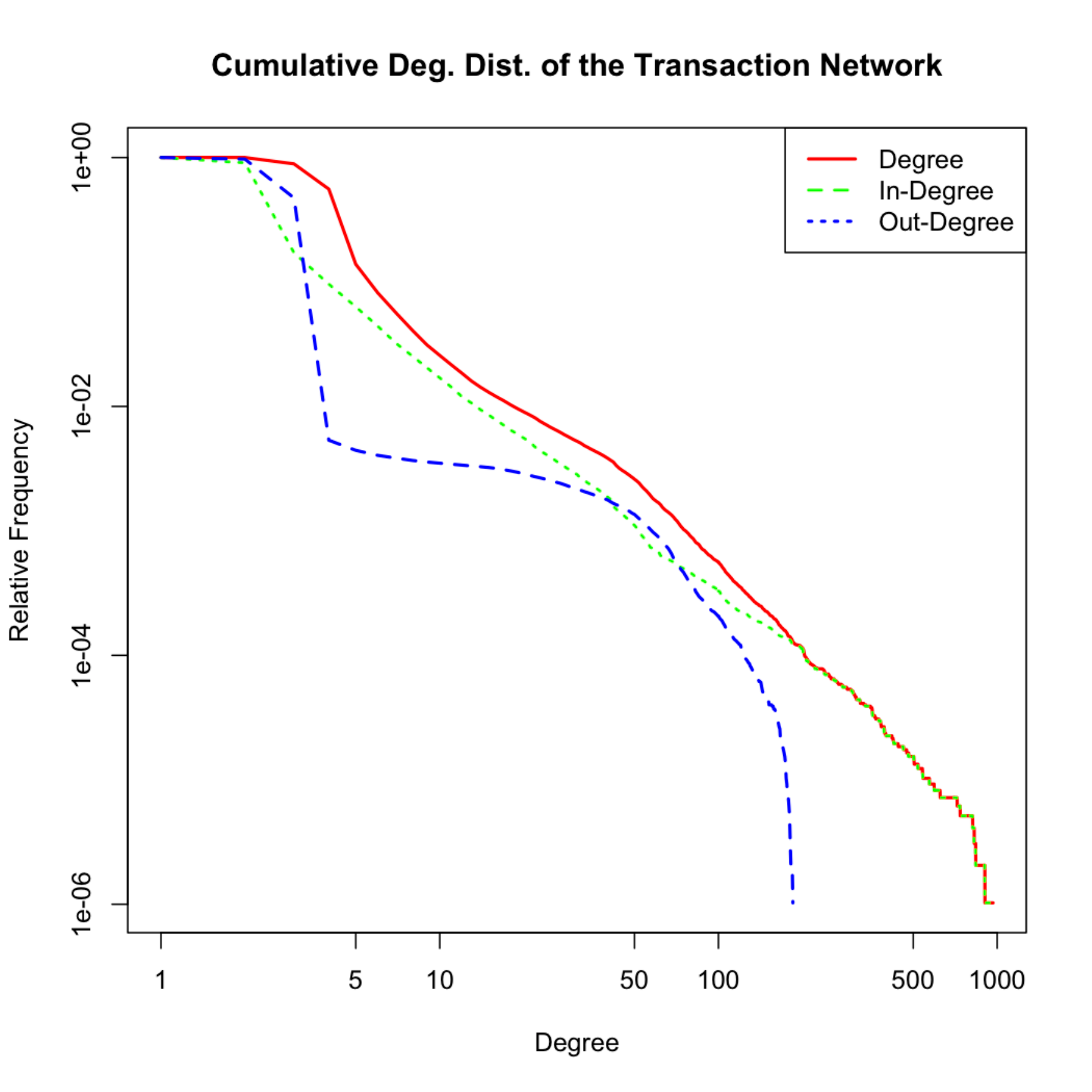}
\label{fig:transaction_network_cum_deg_dist}
}
\subfigure[]{
\includegraphics[width=6.0cm]{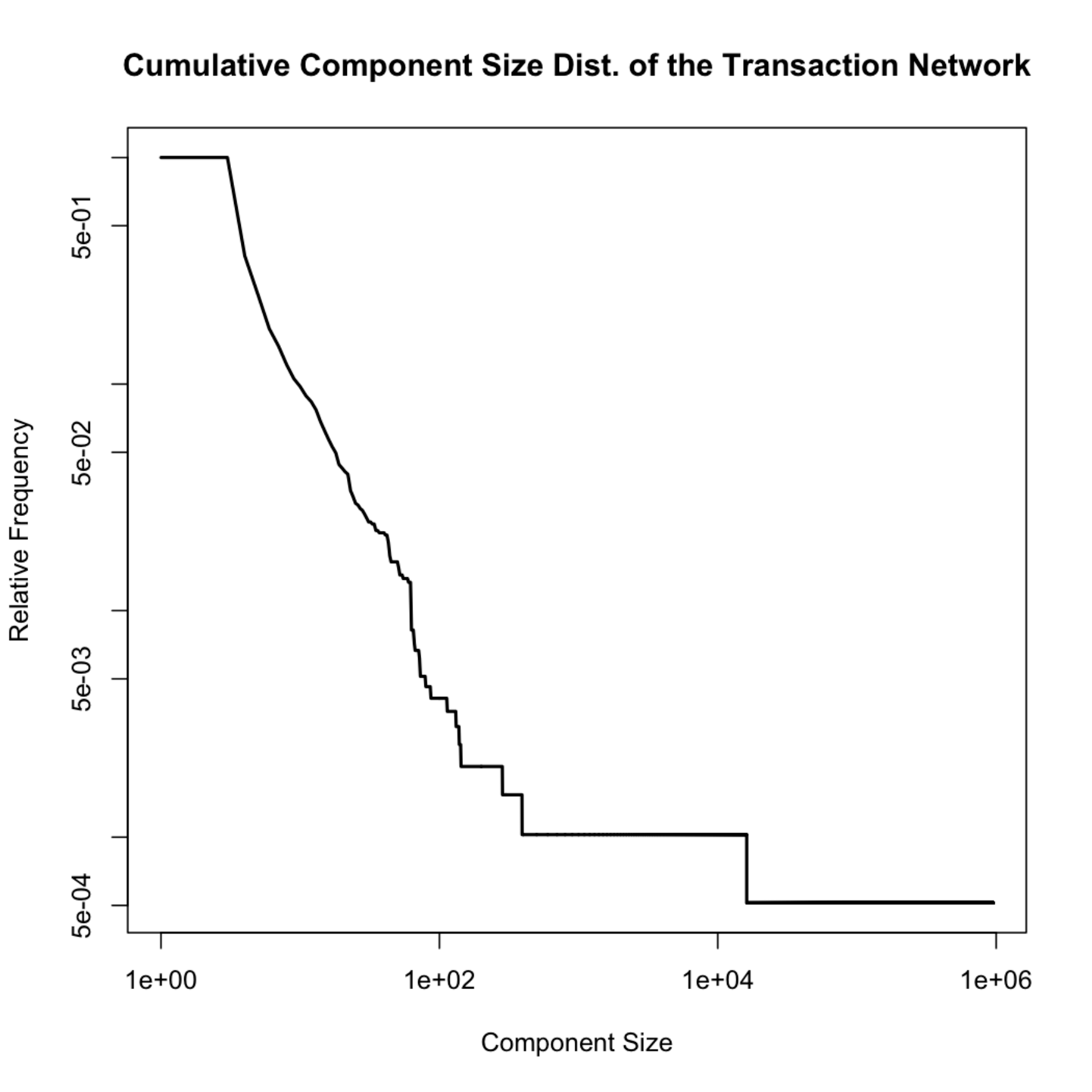}
\label{fig:transaction_network_cum_comp_size_dist}
}
}
\centerline{
\subfigure[]{
\includegraphics[width=4.0cm]{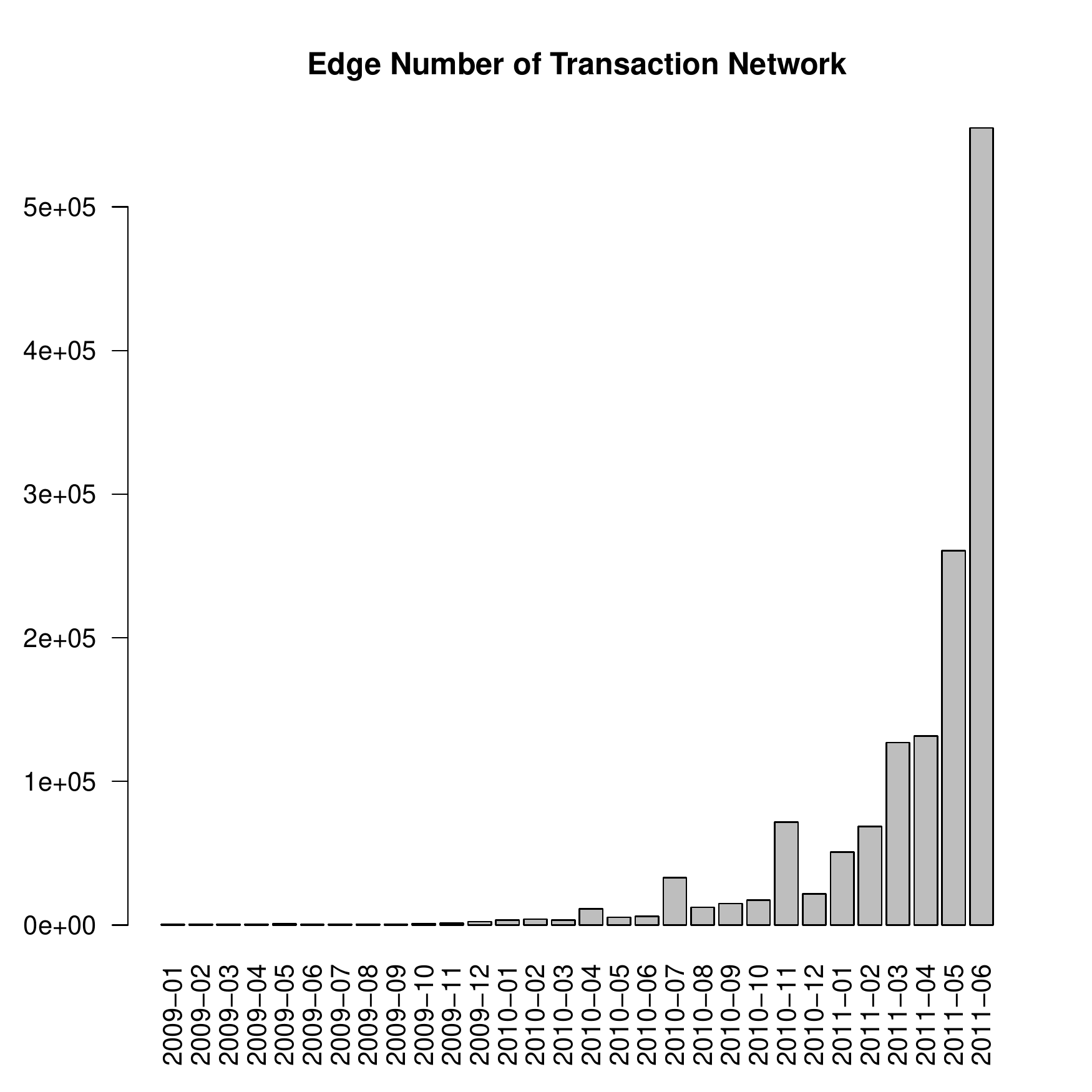}
\label{fig:transaction_network_dyn_edge_number}
}
\subfigure[]{
\includegraphics[width=4.0cm]{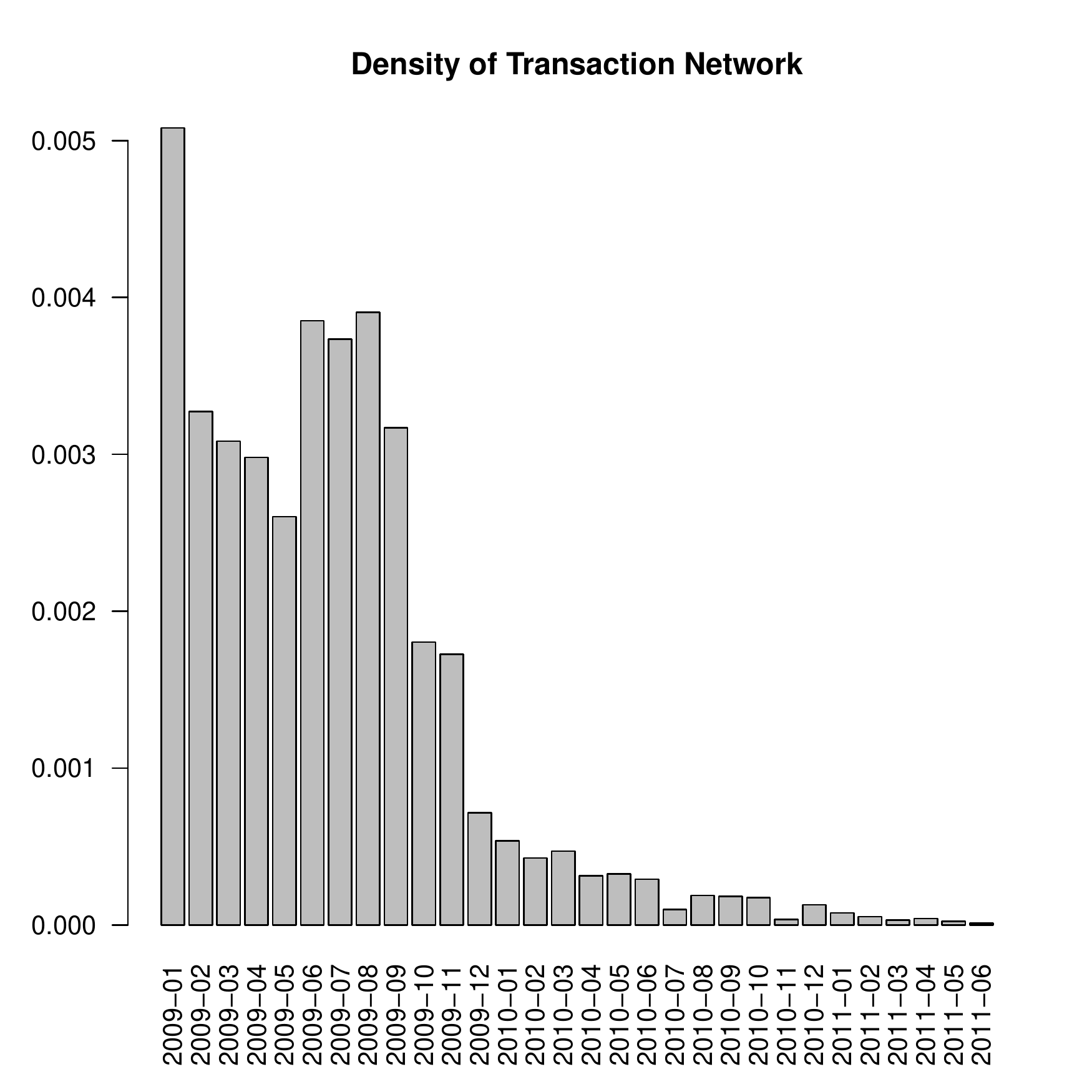}
\label{fig:transaction_network_dyn_density}
}
\subfigure[]{
\includegraphics[width=4.0cm]{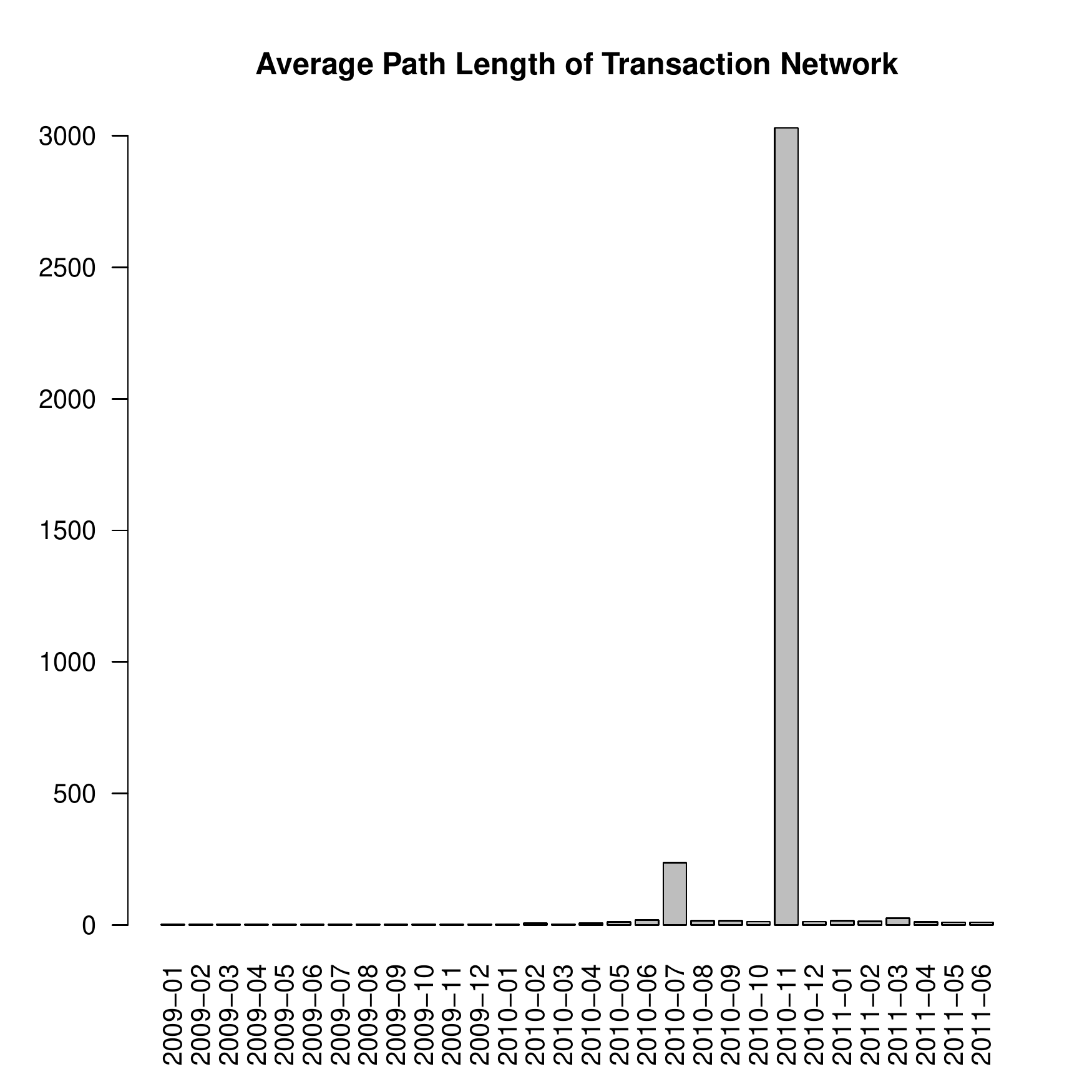}
\label{fig:transaction_network_dyn_avg_path_length}
}
}
\label{fig:transaction_network_analyses}
\caption{For the transaction network: {\bf (a)} A log-log plot of the cumulative degree distributions. {\bf (b)} A log-log plot of the cumulative component size distribution. {\bf (c)} A temporal histogram showing the number of edges per month. {\bf (d)} A temporal histogram showing the density per month. {\bf (e)} A temporal histogram showing the average path length per month.}
\end{figure}

Figure~\ref{fig:transaction_network_cum_deg_dist} shows a log-log plot of
the cumulative degree distributions: the solid red curve is the cumulative
degree distribution (in- and out-degree); the dashed green curve is the
cumulative in-degree distribution; and the dotted blue curve is the
cumulative out-degree distribution. We fitted power-law distributions,
$p(x) \sim x^{-\alpha}$ for $x > x_{min}$, to the three distributions by
estimating the parameters $\alpha$ and $x_{min}$ using a goodness-of-fit
method~\clubpenalty10000\cite{clauset-et-al-09}.
Table~\ref{tbl:transaction_network_deg_dist} shows the estimates along with
the corresponding Kolmogorov--Smirnov goodness-of-fit (GoF) statistics and
p-values. We observe that none of the distributions for which the
empirically-best scaling region is non-trivial have a power-law as a
plausible hypothesis ($p > 0.1$). This is likely due to the fact that there
is no preferential
attachment~\clubpenalty10000\cite{simon-55,barabasi-albert-99}: new
vertices are joined to existing vertices whose corresponding
transactions are not yet fully redeemed.

There are $\numprint{1949}$ (maximal weakly) connected components in the
network. Fig.~\ref{fig:transaction_network_cum_comp_size_dist} shows a log-log
plot of the cumulative component size distribution. There are
$\numprint{948287}$ vertices ($97.31$\%) in the giant component. This
component also contains a giant biconnected component with
$\numprint{716354}$ vertices ($75.54$\% of the vertices in the giant
component).

\begin{table}[ht]
\begin{center}
{\footnotesize
\begin{tabular}{lrrrrrrr}
\textbf{Variable} & $\mathbf{\widetilde{x}}$ & $\mathbf{\bar{x}}$ & $\mathbf{s}$ & $\mathbf{\alpha}$ & $\mathbf{x_{min}}$ & \textbf{GoF} & \textbf{p-val.} \\
\hline
Degree & 3 & 3.20 & 6.20 & 3.24 & 50 & 0.02 & 0.05 \\
In-Degree & 1 & 1.60 & 5.31 & 2.50 & 4 & 0.01 & 0.00 \\
Out-Degree & 1 & 1.60 & 3.17 & 3.50 & 51 & 0.05 & 0.00 \\
\end{tabular}
}
\caption{The degree, in-degree and out-degree distributions of
$\mathcal{T}$.}
\label{tbl:transaction_network_deg_dist}
\end{center}
\end{table}

We also performed a rudimentary dynamic analysis of the network. Figures~\ref{fig:transaction_network_dyn_edge_number}, \ref{fig:transaction_network_dyn_density} and \ref{fig:transaction_network_dyn_avg_path_length} show the edge number, density and average path length of the transaction network on a monthly basis. These measurements are not cumulative. The network's growth and sparsification are evident. We also observe some anomalies in the average path length during July and November 2010.

}{

In the extended version of this paper~\cite{reid-harrigan-11} we produce a log-log plot of the cumulative degree distributions and observe that none of the distributions for which the empirically-best scaling region is non-trivial have a power-law as a plausible hypothesis ($p > 0.1$). We produce a log-log plot of the cumulative component size distribution and observe that there exists considerable cyclic structure. We also performed a rudimentary dynamic analysis of the network considering edge number, density and average path length and highlighted, for example, some anomalies in the average path length during July and November 2010.

}

\subsection{ The User Network }
\label{sec:the_user_network}

The user network $\mathcal{U}$ represents the flow of Bitcoins between
\emph{users} over time. Each vertex represents a user and each directed
edge between a source and a target represents an input-output pair of a
single transaction where the input's public-key belongs to the user
corresponding to the source and the output's public-key belongs to the user
corresponding to the target. Each directed edge also includes a value in
Bitcoins and a timestamp.

We need to perform a preprocessing step before we can construct
$\mathcal{U}$ from our dataset. Suppose $\mathcal{U}$ is, at first,
incomplete in the sense that each vertex represents a single public-key
rather than a user and that each directed edge between a source and a target
represents an input-output pair of a single transaction, where the input's
public-key corresponds to the source and the output's public-key
corresponds to the target. In order to perfect this network, we need to
contract each subset of vertices whose corresponding public-keys belong to
a single user. The difficulty is that public-keys are Bitcoin's mechanism
for ensuring anonymity: `the public can see that someone [identified by a
public-key] is sending an amount to someone else [identified by another
public-key], but without information linking the transaction to
anyone.'~\clubpenalty10000\cite{nakamoto-08}. In fact, it is considered
good practice for a payee to generate a new public-private key-pair for
every transaction to keep transactions from being linked to a common owner.
Therefore, it is impossible to completely perfect the network using our
dataset alone. However, as noted by Nakamoto~\clubpenalty10000\cite{nakamoto-08},

\begin{quote}
``Some linking is still unavoidable with multi-input transactions, which
necessarily reveal that their inputs were owned by the same owner. The risk
is that if the owner of a key is revealed, linking could reveal other
transactions that belonged to the same owner.''
\end{quote}

We will use this property of transactions with multiple inputs to contract
subsets of vertices in the incomplete network. We construct an ancillary
network in which each vertex represents a public-key.
We connect pairs of vertices with undirected edges, where each edge joins a
pair of public keys that are both inputs to the same transaction and are thus
controlled by the same user. 
From our dataset, this ancillary network has $\numprint{1253054}$ vertices (unique public-keys) and
$\numprint{4929950}$ edges. More importantly, it has $\numprint{86641}$
non-trivial maximal connected components.
Each maximal connected component in this graph corresponds to a user, and each component's
constituent vertices correspond to that user's public-keys.

\ifthenelse {\boolean{longVersion}} {

\begin{figure}[!h]
\centerline{
\subfigure[]{
\includegraphics[width=6.0cm]{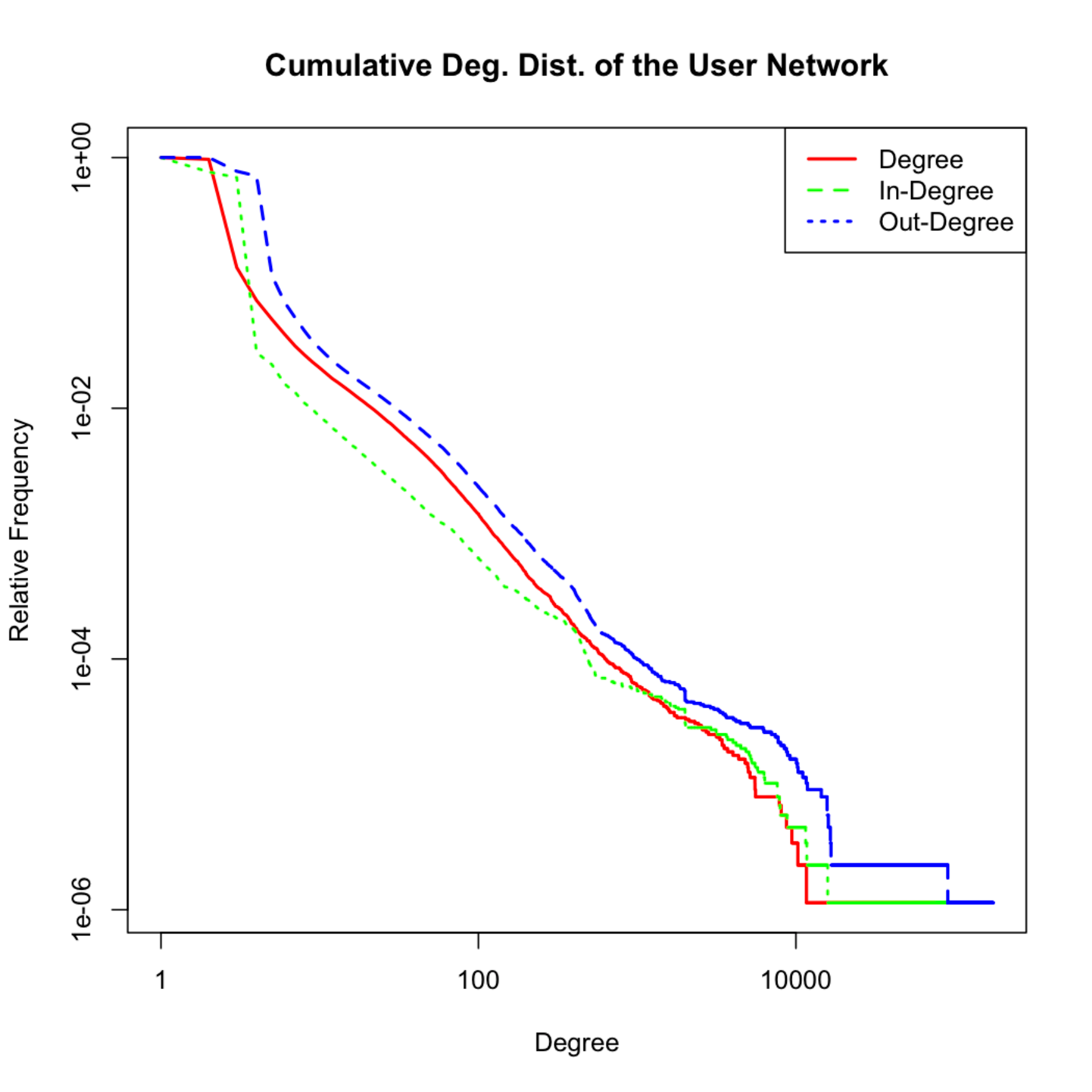}
\label{fig:user_network_cum_deg_dist}
}
\subfigure[]{
\includegraphics[width=6.0cm]{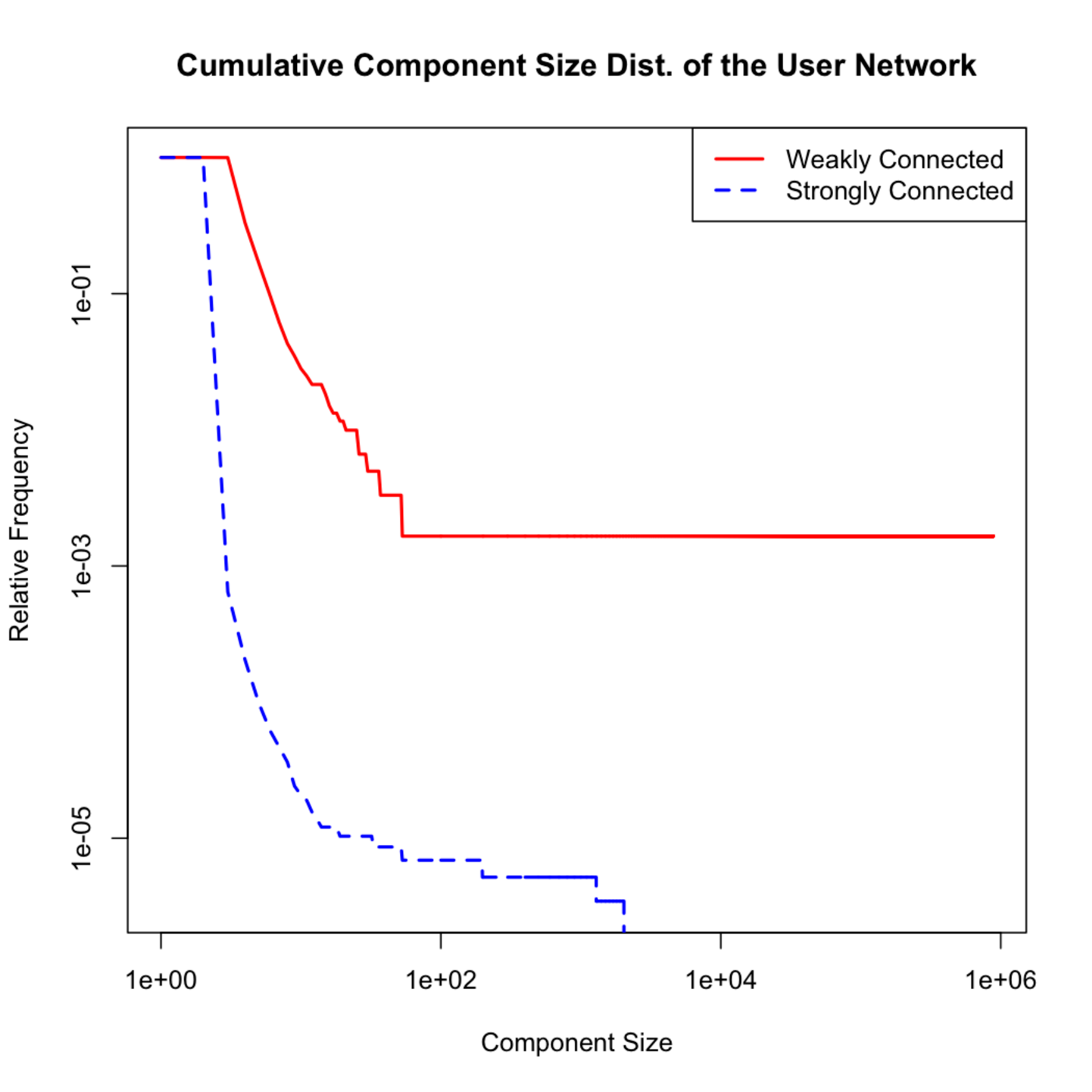}
\label{fig:user_network_cum_comp_size_dist}
}
}
\centerline{
\subfigure[]{
\includegraphics[width=4.0cm]{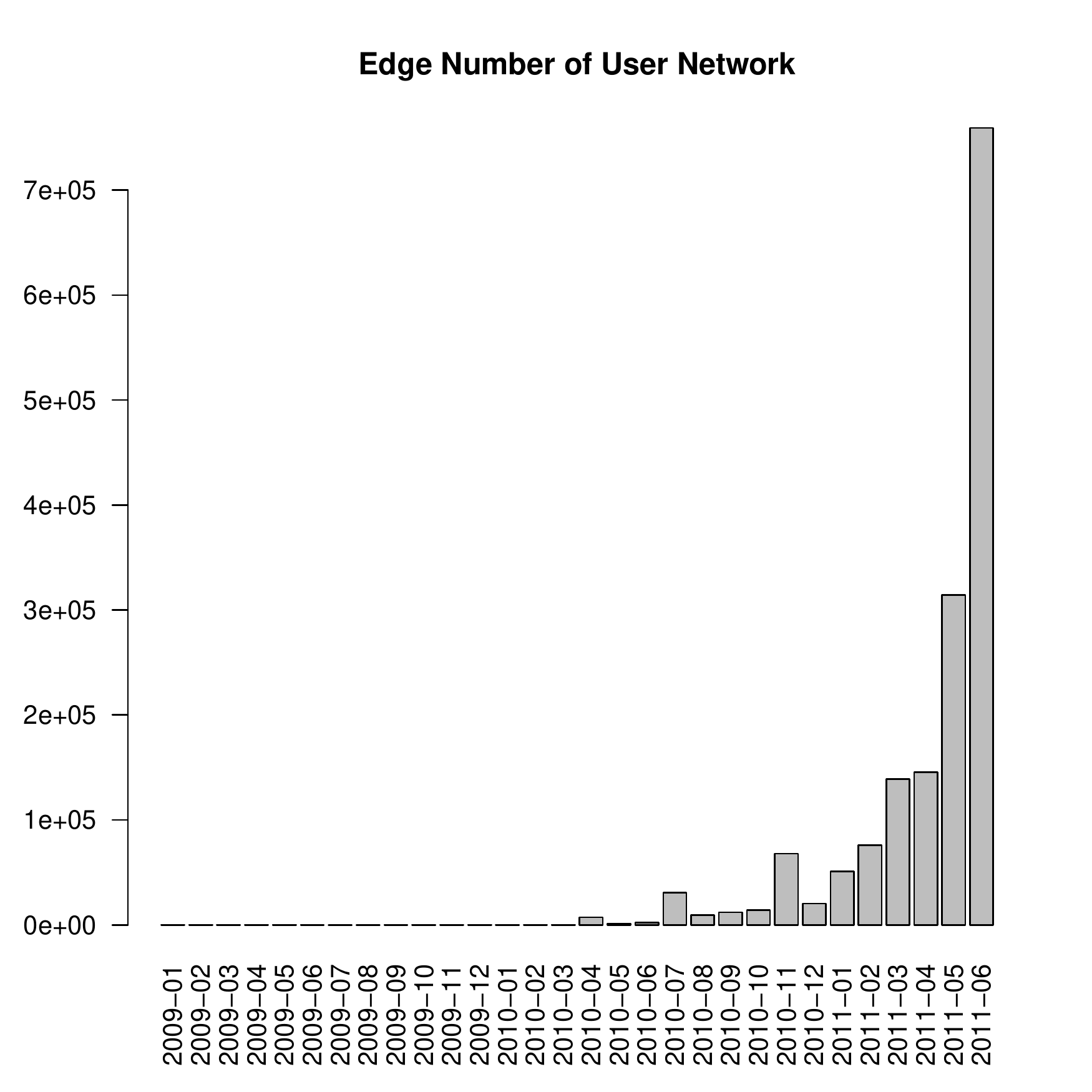}
\label{fig:user_network_dyn_edge_number}
}
\subfigure[]{
\includegraphics[width=4.0cm]{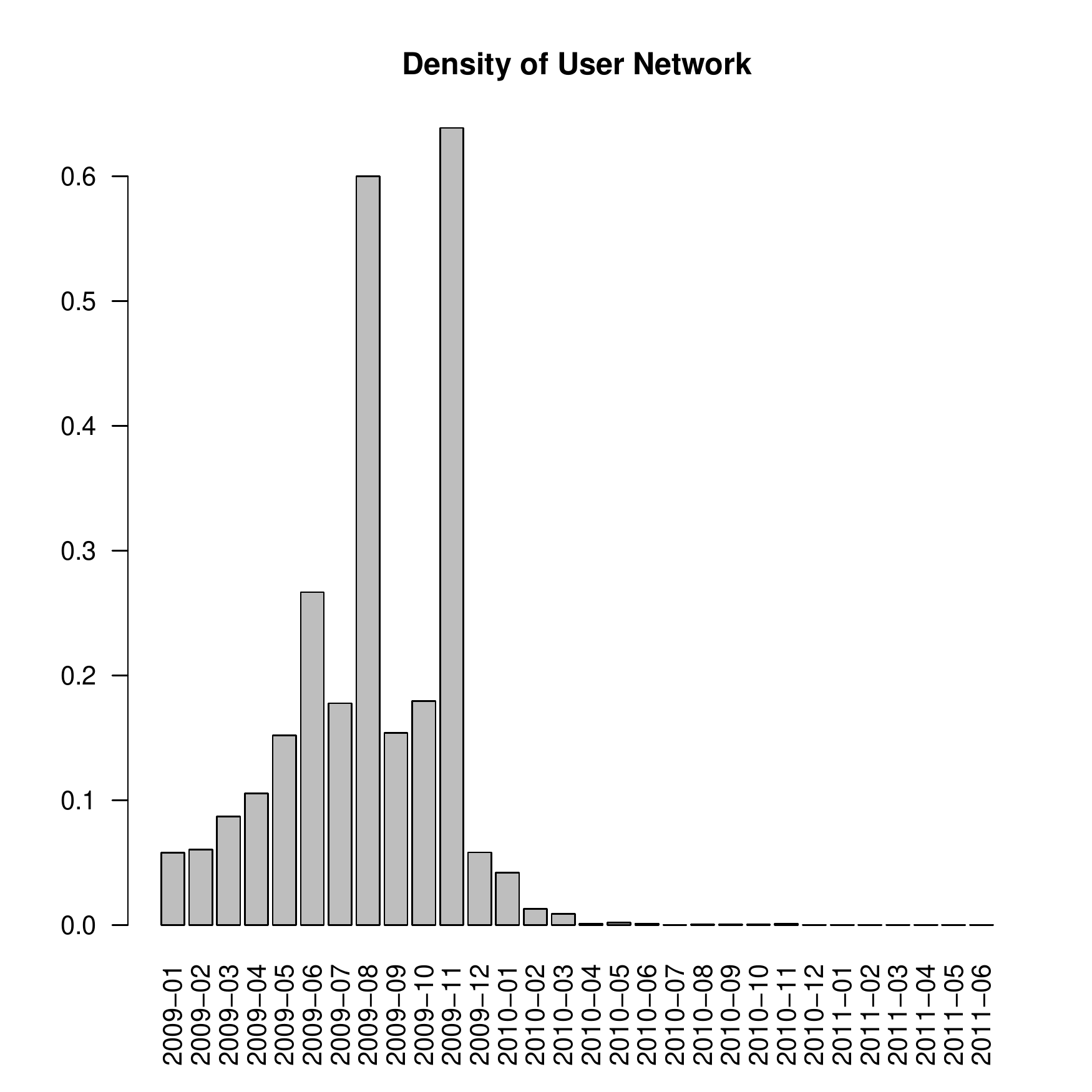}
\label{fig:user_network_dyn_density}
}
\subfigure[]{
\includegraphics[width=4.0cm]{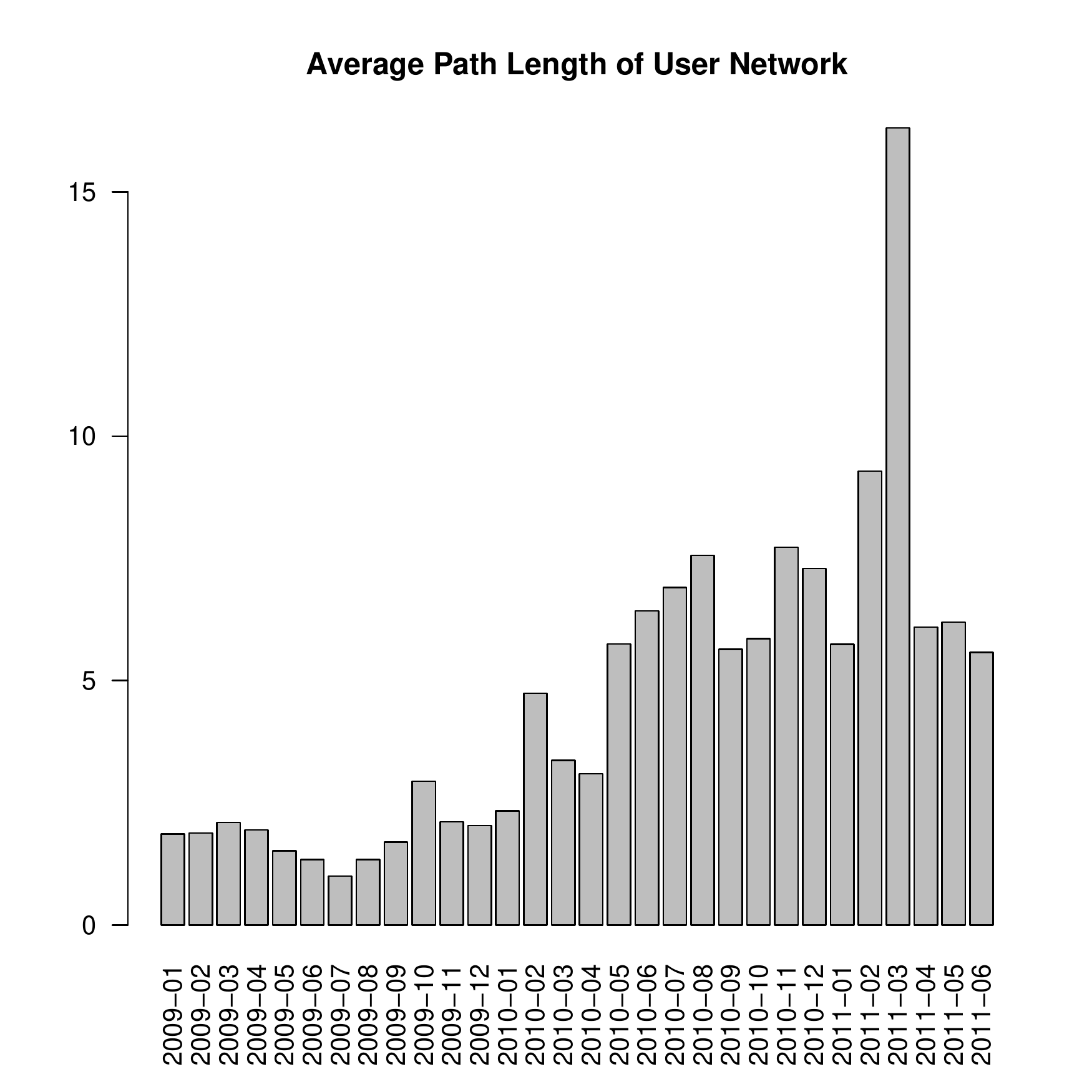}
\label{fig:user_network_dyn_avg_path_length}
}
}
\label{fig:user_network_analyses}
\caption{For the user network: {\bf (a)} A log-log plot of the cumulative degree distributions. {\bf (b)} A log-log plot of the cumulative component size distribution. {\bf (c)} A temporal histogram showing the number of edges per month. {\bf (d)} A temporal histogram showing the density per month. {\bf (e)} A temporal histogram showing the average path length per month.}
\end{figure}

}

\begin{figure}\centerline{\includegraphics[width=7.5cm]{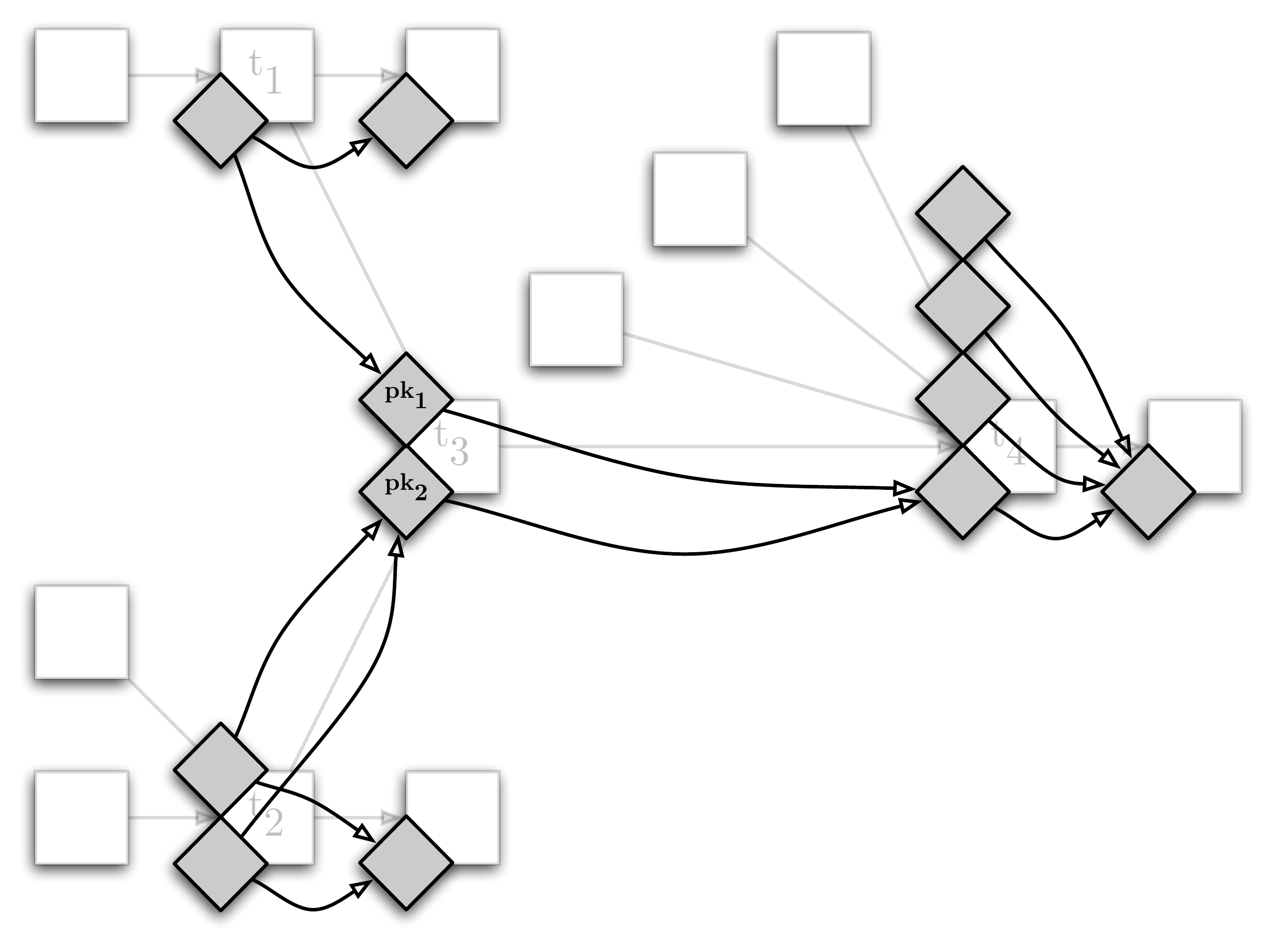}}\caption{An
example sub-network from the incomplete network. Each diamond vertex represents a
public-key and each directed edge between diamond vertices represents a
flow of Bitcoins from one public-key to
another.}\label{fig:public_key_network_example}\end{figure}

\begin{figure}\centerline{\includegraphics[width=7.5cm]{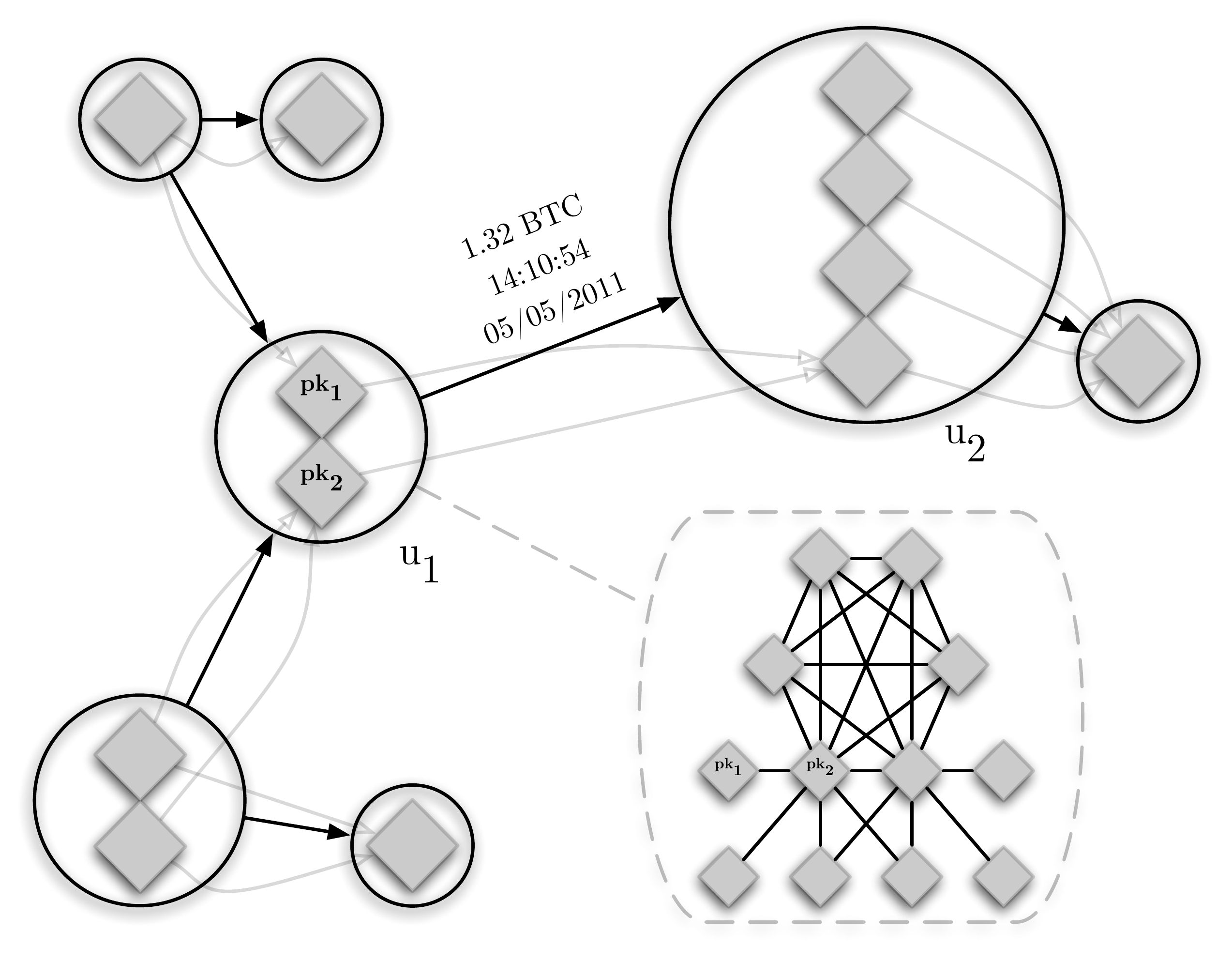}}\caption{An
example sub-network from the user network. Each circular vertex represents a user and
each directed edge between circular vertices represents a flow of Bitcoins
from one user to another. The maximal connected component from the ancillary network
that corresponds to the vertex $u_1$ is shown within the dashed grey box.}\label{fig:user_network_example}

\vspace{-10px}
\end{figure}

Figure~\ref{fig:public_key_network_example} shows an example
sub-network of the incomplete network overlaid onto the example sub-network of
$\mathcal{T}$ from
Fig.~\ref{fig:transaction_network_example}. The outputs
of $t_1$ and $t_2$ that were eventually redeemed by $t_3$ were sent to a user whose
public-key was $pk_1$ and a user whose public-key was $pk_2$ respectively.
Figure~\ref{fig:user_network_example} shows an example sub-network
of the user network overlaid onto the example sub-network of the incomplete network
from Fig.~\ref{fig:public_key_network_example}. $pk_1$
and $pk_2$ are contracted into a single vertex $u_1$ since they correspond
to a pair inputs of a single transaction. In other words, they are in the
same maximal connected component of the ancillary network (see the vertices
representing $pk_1$ and $pk_2$ in the dashed grey box in
Fig.~\ref{fig:user_network_example}). A single user owns both public-keys.
We note that the maximal connected component in this case
is not simply a clique; it has a diameter of four indicating that there are
at least two public-keys belonging to that same user that are connected
indirectly via three transactions. The sixteen inputs to transaction $t_4$
result in the contraction of a further sixteen public-keys into a single vertex
$u_2$. The value and timestamp of the flow of Bitcoins from $u_1$ to
$u_2$ is derived from the transaction network.

After the preprocessing step, $\mathcal{U}$ has $\numprint{881678}$
vertices ($\numprint{86641}$ non-trivial maximal connected components and
$\numprint{795037}$ isolated vertices in the ancillary network) and
$\numprint{1961636}$ directed edges. The network is still incomplete. We
have not contracted all possible vertices but it will suffice for our
present analysis. Unlike $\mathcal{T}$, $\mathcal{U}$ has multi-edges,
loops and directed cycles.

\ifthenelse {\boolean{longVersion}} {

Figure~\ref{fig:user_network_cum_deg_dist} shows a log-log plot of the
network's cumulative degree distributions. We fitted power-law
distributions to the three distributions and calculated their
goodness-of-fit and statistical significance as in the previous section.
Table~\ref{tbl:user_network_deg_dist} shows the results. We observe that
none of the distributions have a power-law as a plausible hypothesis.

There are $\numprint{604}$ (maximal) weakly connected components and
$\numprint{579355}$ (maximal) strongly connected components in the network;
Fig.~\ref{fig:user_network_cum_comp_size_dist} shows a log-log plot of the
cumulative component size distribution for both variations. There are
$\numprint{879859}$ vertices ($99.79$\%) in the giant weakly connected
component. This component also contains a giant weakly biconnected
component with $\numprint{652892}$ vertices ($74.20$\% of the vertices in
the giant component).

\begin{table}[ht]
\begin{center}
{\footnotesize
\begin{tabular}{lrrrrrrr}
\textbf{Variable} & $\mathbf{\widetilde{x}}$ & $\mathbf{\bar{x}}$ & $\mathbf{s}$ & $\mathbf{\alpha}$ & $\mathbf{x_{min}}$ & \textbf{GoF} & \textbf{p-val.} \\
\hline
Degree & 3 & 4.45 & 218.10 & 2.38 & 66 & 0.02 & 0.00 \\
In-Degree & 1 & 2.22 & 86.40 & 2.45 & 57 & 0.05 & 0.00 \\
Out-Degree & 2 & 2.22 & 183.91 & 2.03 & 10 & 0.22 & 0.00 \\
\end{tabular}
}
\caption{The degree, in-degree and out-degree distributions of
$\mathcal{U}$.}
\label{tbl:user_network_deg_dist}
\end{center}
\end{table}

Our dynamic analysis of the user network mirrors that of the transaction network in the previous subsection. Figures~\ref{fig:user_network_dyn_edge_number}, \ref{fig:user_network_dyn_density} and \ref{fig:user_network_dyn_avg_path_length} show the edge number, density and average path length of the user network on a monthly basis. These measurements are not cumulative. The network's growth and sparsification are evident. We note that even though our dynamic analysis of the user network is on a monthly basis, the preprocessing step is performed using the ancillary network of the entire incomplete network. This enables us to resolve public-keys to a single user irrespective of the month in which the linking transactions occur.

} {

In the extended version of this paper~\cite{reid-harrigan-11} we produce a log-log plot of the cumulative degree distributions and observe that none of the distributions have a power-law as a plausible hypothesis ($p > 0.1$). We produce a log-log plot of the cumulative component size distribution and observe that there exists considerable cyclic structure. We also performed a rudimentary dynamic analysis of the network considering edge number, density and average path length.

}

The contraction of public-keys into users, while incomplete, generates a network that is in many ways a proxy for the social network of Bitcoin users. The edges represent financial transactions between pairs of users. It may be possible to identify, for example, communities, central users and hoarders within this social network.

\section{Anonymity Analysis}
\label{sec:anonymity_analysis}

Prior to performing the analyses above, we expected the user network to be largely composed of trees representing Bitcoin flows between one-time public-keys that were not linked with other public-keys. However, our analyses reveal that the user network has considerable cyclic structure. We now consider the implications of this structure, coupled with other aspects of the Bitcoin system, for anonymity.

There are several ways in which the user network can be used to deduce information about Bitcoin users. We can use global network properties, such as degree distribution, to identify outliers. We can use local network properties to examine the context in which a user operates by observing the users with which he or she interacts with either directly or indirectly. The dynamic nature of the user network also enables us to perform flow and temporal analyses. We can examine the significant Bitcoin flows between groups of users over time. We will now discuss each of these possibilities in more detail and provide a case study to demonstrate their use in practice.

\subsection{Integrating Off-Network Information}
\label{sec:integrating_off_network_information}

There is no user directory for the Bitcoin system. However, we can attempt to build a partial user directory associating Bitcoin users (and their known public-keys) with off-network information. If we can make sufficient associations and combine them with the network structures above, a potentially serious threat to anonymity emerges.

Many organizations and services such as on-line stores that accept Bitcoins, exchanges, laundry services and mixers have access to identifying information regarding their users, \eg e-mail addresses, shipping addresses, credit card and bank account details, IP addresses, etc. If any of this information is publicly available, or accessible by, say, law enforcement agencies, then the identities of users involved in related transactions may also be at risk. To illustrate this point, we consider a number of publicly available data sources and integrate their information with the user network.

\ifthenelse {\boolean{longVersion}} {
\subsubsection{The Bitcoin Faucet}
}

The Bitcoin Faucet\footnote{http://freebitcoins.appspot.com} is a website where users can donate Bitcoins to be redistributed in small amounts to other users. In order to prevent abuse of this service, a history of recent give-aways are published along with the IP addresses of the recipients. When the Bitcoin Faucet does not batch the re-distribution, it is possible to associate the IP addresses with the recipient's public-keys. This page can be scraped over time to produce a time-stamped mapping of IP addresses to users.

We found that the public-keys associated with many of the IP addresses that received Bitcoins were contracted with other public-keys in the ancillary network, thus revealing IP addresses that are somehow related to previous transactions.
\ifthenelse {\boolean{longVersion}} {
Fig.~\ref{fig:faucetIPMap} shows a map of geolocated IP addresses belonging to users who received Bitcoins over a period of one week. Fig.~\ref{fig:faucetIPNetwork} overlays the user network onto a sample of those users. An edge between two geolocated IP addresses indicates that the corresponding users are linked by an undirected path of length at most three in the user network; the path must not contain the vertex representing the Bitcoin Faucet itself.

These figures serve as a proof-of-concept from a small publicly available data source. We note that large centralized Bitcoin service providers are capable of producing much more detailed maps.

\begin{figure*}[ht]
\centerline{
\subfigure[]{
\includegraphics[width=6cm]{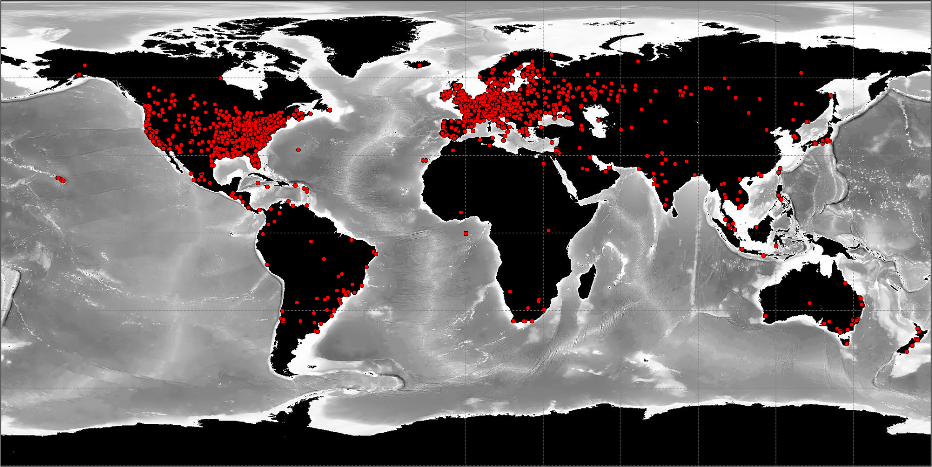}
\label{fig:faucetIPMap}
}
\subfigure[]{
\includegraphics[width=6cm]{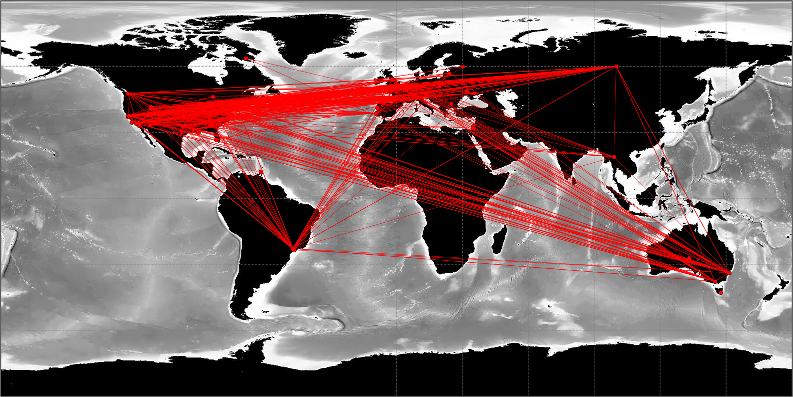}
\label{fig:faucetIPNetwork}
}
}
\label{fig:MapNetworks}
\caption{We can use the Bitcoin Faucet to map users to geolocated IP addresses. {\bf(a)} A map of geolocated IP addresses associated with users receiving Bitcoins from the Bitcoin Faucet during a one week period. {\bf(b)} A map of a sample of the geolocated IP addresses in \subref{fig:faucetIPMap} connected by edges where the corresponding users are connected by a path of length at most three in the user network that does not include the vertex representing the Bitcoin Faucet.}
\end{figure*}

}

\ifthenelse {\boolean{longVersion}} {
\subsubsection{Voluntary Disclosures}
}

Another source of identifying information is the voluntary disclosure of public-keys by users, for example, when posting to the Bitcoin forums\footnote{http://forum.bitcoin.org}. Bitcoin public-keys are typically represented as strings approximately thirty-three characters in length and starting with the digit one. They are indexed very well by popular search engines. We identified many high-degree vertices with external information using a search engine alone. We scraped the Bitcoin Forums where users frequently attach a public-key to their signatures. We also gathered public-keys from Twitter streams and user-generated public directories. It is important to note that in many cases we are able to resolve the `public' public-keys with other public-keys belonging to the same user using the ancillary network. We also note that large centralized Bitcoin service providers can do the same with their user information.

\subsection{TCP/IP Layer Information}
\label{sec:tcp_ip_layer_information}
Security researcher Dan Kaminsky has performed an analysis of the Bitcoin system, investigating identity leakage at the TCP/IP layer.
He found that by opening a connection to all public peers in the network at once, he could map IP addresses to Bitcoin public-keys, working from the assumption that ``the first node to inform you of a transaction is the source of it\ldots [this is] more or less true, and absolutely over time''~\cite{kaminsky}.
Using this approach it is possible to map public-keys to IP addresses unless users are using an anonymising proxy technology such as TOR.

\subsection{Egocentric Analysis and Visualization}
\label{sec:egocentric_analysis_and_visualization}

\begin{figure}[ht]
\center{
\subfigure[The receipts and payments to and from WikiLeaks' public-key over time.]{
\includegraphics[width=50mm]{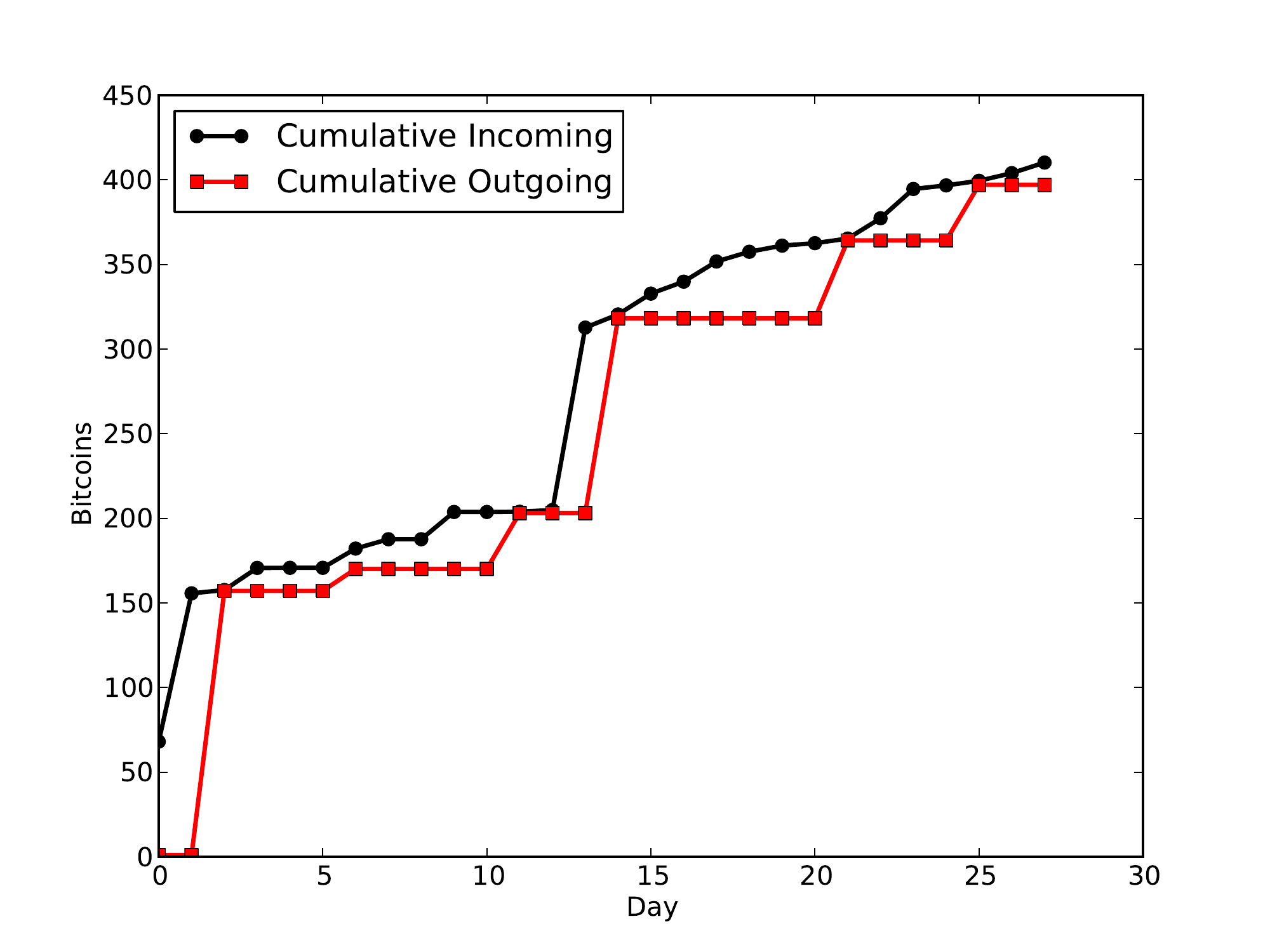}
\label{fig:wikileaksCumulInOut}
}
\hspace{5mm}
\subfigure[The number of transactions involving WikiLeaks' public-key over time.]{
\includegraphics[width=50mm]{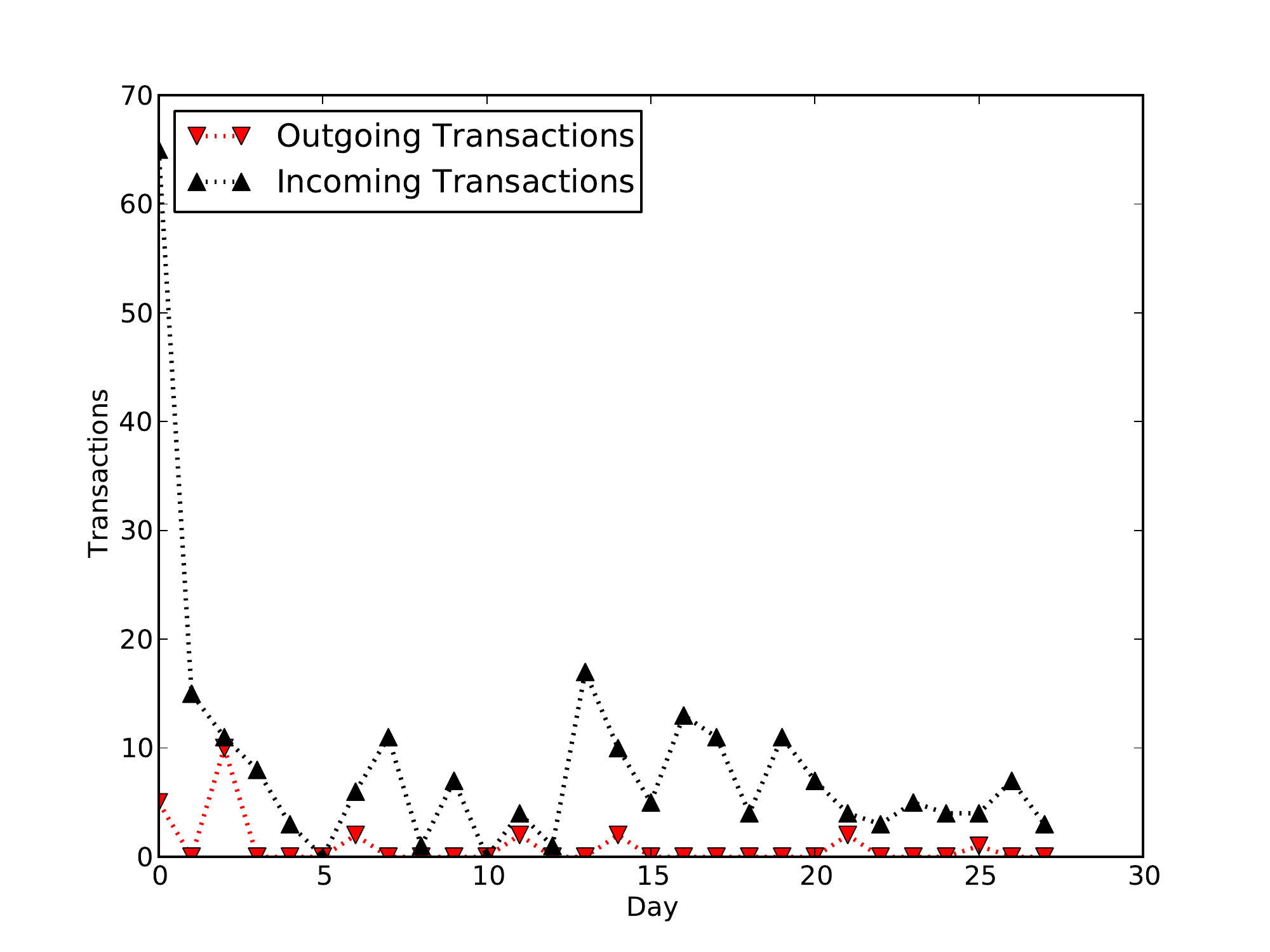}
\label{fig:wikileaksTransInOut}
}
}
\center{
\subfigure[The receipts and payments to and from the creator of a popular Bitcoin trading website aggregated over a number of public-keys.]{
\includegraphics[width=50mm]{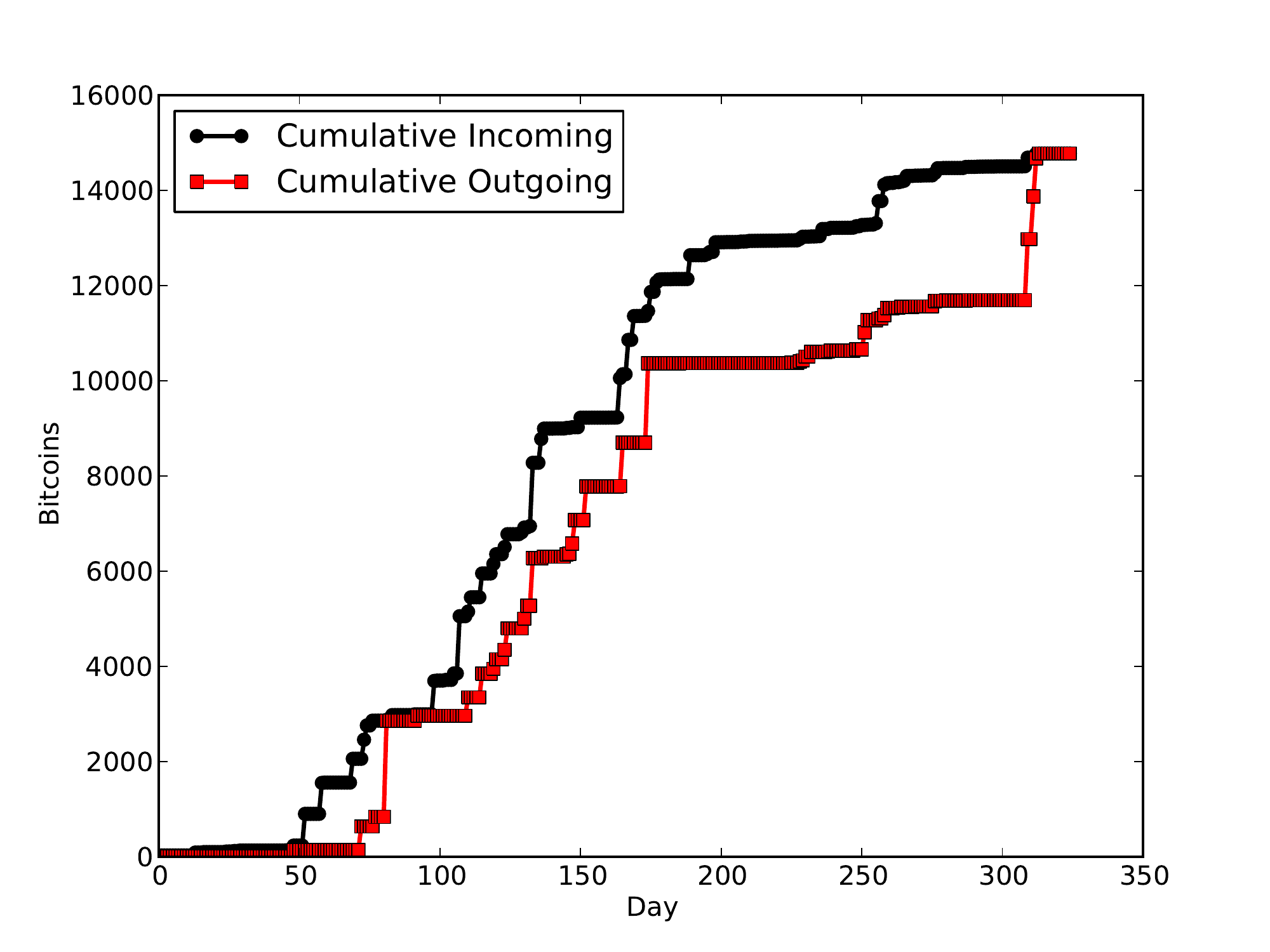}
\label{fig:bitcoinotcCumulInOut}
}
}
\label{fig:detailedStats}
\caption{We can plot cumulative receipts and payments to and from Bitcoin public-keys and users.}
\end{figure}

There are severals pieces of information we can directly derive from the user network regarding a particular user. We can compute the balance held by a single public-key. We can also aggregate the balances belonging to public-keys that are controlled by a particular user. For example, Fig.~\ref{fig:wikileaksCumulInOut} and Fig.~\ref{fig:wikileaksTransInOut} show the receipts and payments to and from WikiLeaks' public-key in terms of Bitcoins and the number of transactions respectively. The donations are relatively small and are forwarded to other public-keys periodically. There was also a noticeable spike in donations when the facility was first announced. Figure~\ref{fig:bitcoinotcCumulInOut} shows the receipts and payments to and from the creator of a popular Bitcoin trading website aggregated over a number of public-keys that are linked through the ancillary network.

\begin{figure}[!htb]
\begin{center}
        \includegraphics[width=70mm]{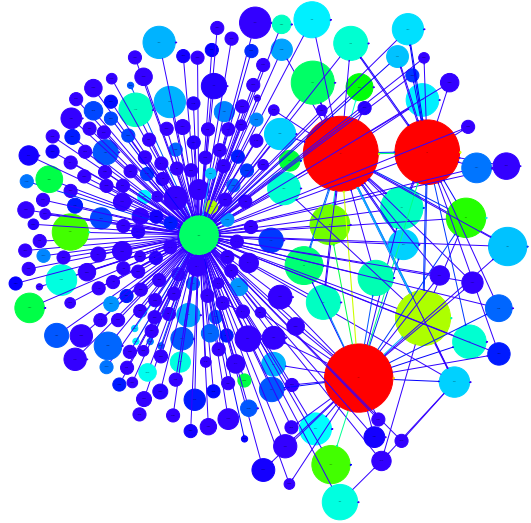}
\end{center}
\caption{An egocentric visualization of the vertex representing WikiLeaks' public-key in the incomplete user network. The size of a vertex corresponds to its degree in the entire incomplete user network. The color denotes the volume of Bitcoins -- warmer colors have larger volumes flowing through them. The three largest red vertices represent a Bitcoin mining pool, a centralized Bitcoin wallet service, and an unknown entity.}
\label{fig:wikileaksEgo}
\end{figure}

An important advantage of deriving network structures from the Bitcoin transaction history is our ability to use network visualization and analysis tools to investigate the flow of Bitcoins. For example, Fig.~\ref{fig:wikileaksEgo} shows the network structure surrounding the WikiLeaks' public-key in the incomplete user network. Our tools resolve several of the vertices with identifying information gathered in Sect.~\ref{sec:integrating_off_network_information}. These users can be linked either directly or indirectly to their donations.

\subsection{Context Discovery}
\label{sec:context_discovery}

\begin{figure}[!htb]
\begin{center}
        \includegraphics[width=70mm]{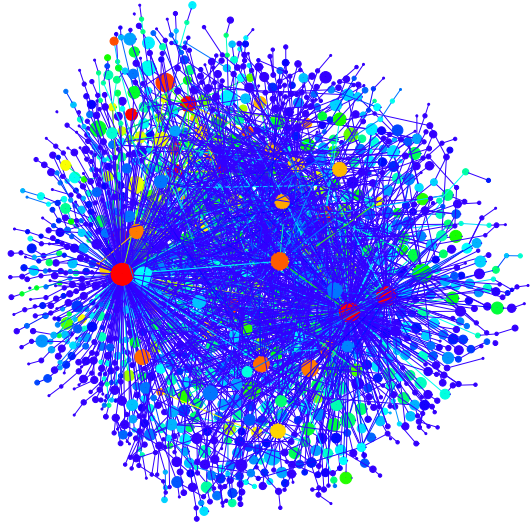}
\end{center}
\caption{A visualisation of all users identified in Sect.~\ref{sec:integrating_off_network_information} and all shortest paths between the vertices representing those users and the vertex representing the MyBitcoin service in the user network.}
\label{fig:visualisationAllDB}
\end{figure}

Given a number of public-keys or users of interest, we can use network structure and context to better understand the flow of Bitcoins between them. For example, we can examine all shortest paths between a set of vertices or consider the maximum number of Bitcoins that can flow from a source to a destination given the transactions and their `capacities' in an interesting time-window. For example, Fig.~\ref{fig:visualisationAllDB} shows all shortest paths between the vertices representing the users we identified using off-network information in Sect.~\ref{sec:integrating_off_network_information} and the vertex that represents the MyBitcoin service\footnote{http://www.mybitcoin.com} in the user network. We can identify more than 60\% of the users in this visualization and deduce many direct and indirect relationships between them.

\emph{Case Study -- Part I:} We analyse an alleged theft of \numprint{25000} BTC reported in the Bitcoin Forums\footnote{http://forum.bitcoin.org/index.php?topic=16457.0} by a user known as \texttt{allinvain}. The victim reported that a large portion of his Bitcoins were sent to $pk_{red}$\footnote{1KPTdMb6p7H3YCwsyFqrEmKGmsHqe1Q3jg} on 13/06/2011 at 16:52:23 UTC. The theft occurred shortly after somebody broke into the victim's Slush pool account\footnote{http://mining.bitcoin.cz} and changed the payout address to $pk_{blue}$.\footnote{15iUDqk6nLmav3B1xUHPQivDpfMruVsu9f}. The Bitcoins rightfully belonged to $pk_{green}$.\footnote{1J18yk7D353z3gRVcdbS7PV5Q8h5w6oWWG} At the time of theft, the stolen Bitcoins had a market value of approximately half a million U.S. dollars. We chose this case study to illustrate the potential risks to the anonymity of a user (the thief) who has good reason to remain anonymous.

\begin{figure}[!htb]
\begin{center}
        \includegraphics[width=80mm]{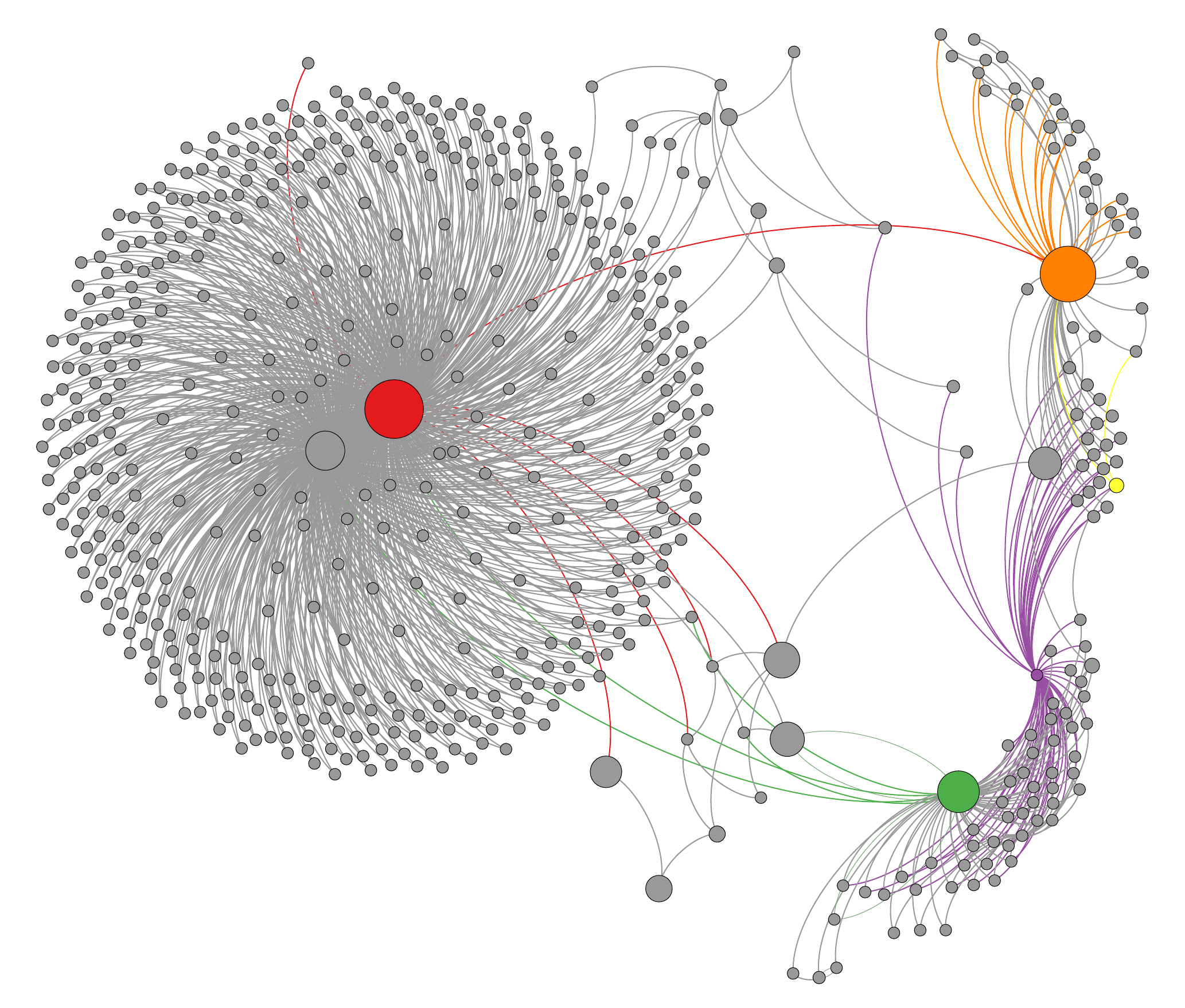}
\end{center}
\caption{An egocentric visualization of the thief in the incomplete user network. For this visualization, vertices are colored according to the text, edges are colored according to the color of their sources and the size of each vertex is proportional to its edge-betweenness within the egocentric network.}
\label{fig:thief_public_key_2_steps_with_extras}
\end{figure}

We consider the incomplete user network before any contractions. We restrict ourselves to the egocentric network surrounding the thief: we include every vertex that is reachable by a path of length at most two ignoring directionality and all edges induced by these vertices. We also remove all loops, multiple edges and edges that are not contained in some biconnected component to avoid clutter. In Fig.~\ref{fig:thief_public_key_2_steps_with_extras}, the red vertex represents the thief who owns the public-key $pk_{red}$ and the green vertex represents the victim who owns the public-key $pk_{green}$. The theft is represented by a green edge joining the victim to the thief.

\begin{figure}[!htb]
\begin{center}
        \includegraphics[width=80mm]{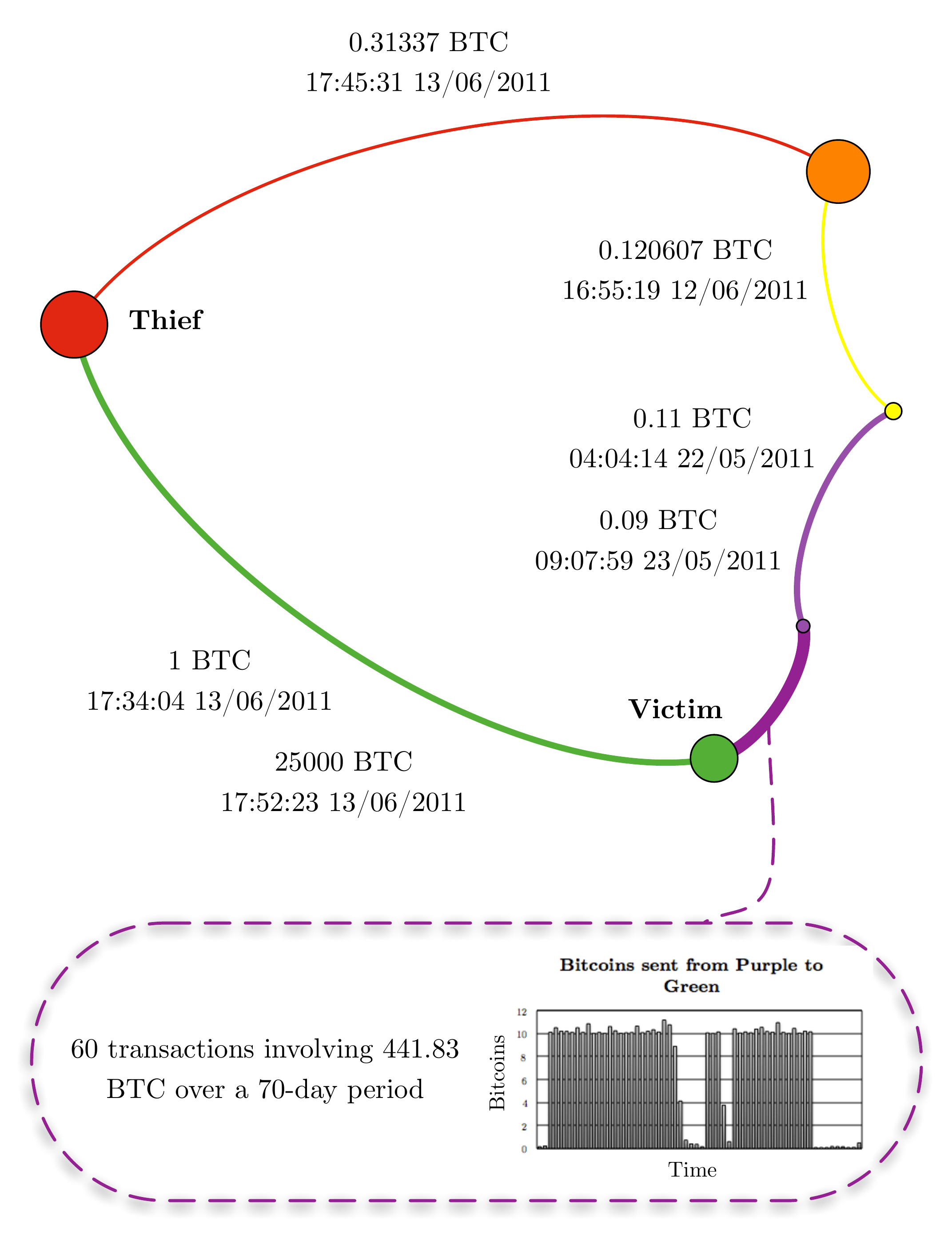}
\end{center}
\caption{An interesting sub-network induced by the thief, the victim and three other vertices. The notation is the same as in Fig.~\ref{fig:thief_public_key_2_steps_with_extras}.}
\label{fig:thief_public_key_2_steps_only_extras_annotated}
\end{figure}

Interestingly, the victim and the thief are joined by paths (ignoring directionality) other than the green edge representing the theft. For example, consider the sub-network shown in Fig.~\ref{fig:thief_public_key_2_steps_only_extras_annotated} induced by the red, green, purple, yellow and orange vertices. This sub-network is a cycle. We contract all vertices whose corresponding public-keys belong to the same user. This allows us to attach values in Bitcoins and timestamps to the directed edges. We can make a number of observations. Firstly, we note that the theft of \numprint{25000} BTC was preceded by a smaller theft of 1 BTC. This was later reported by the victim in the Bitcoin forums. Secondly, using off-network data, we have identified some of the other colored vertices: the purple vertex represents the main Slush pool account and the orange vertex represents the computer hacker group known as LulzSec.\footnote{http://twitter.com/LulzSec/status/76388576832651265}
\ifthenelse {\boolean{longVersion}} {
We note that there has been at least one attempt to associate the thief with LulzSec\footnote{http://pastebin.com/88nGp508}. This was a fake; it was created after the theft. However, the identification of the orange vertex with LulzSec is genuine and was established before the theft.
}

We observe that the thief sent 0.31337 BTC to LulzSec shortly after the theft but we cannot otherwise associate him with the group. The main Slush pool account sent a total of 441.83 BTC to the victim over a 70-day period. It also sent a total of 0.2 BTC to the yellow vertex over a two day period. One day before the theft, the yellow vertex also sent 0.120607 BTC to LulzSec.

The yellow vertex represents a user who is the owner of at least five public-keys.\ifthenelse {\boolean{longVersion}} {\footnote{1MUpbAY7rjWxvLtUwLkARViqSdzypMgVW4\\13tst9ukW294Q7f6zRJr3VmLq6zp1C68EK\\1DcQvXMD87MaYcFZqHzDZyH3sAv8R5hMZe\\1AEW9ToWWwKoLFYSsLkPqDyHeS2feDVsVZ\\1EWASKF9DLUCgEFqfgrNaHzp3q4oEgjTsF}
}

Like the victim, he is a member of the Slush pool, and like the thief, he is a one-time donator to LulzSec. This donation, the day before the theft, is his last known activity using these public-keys.

\subsection{Flow and Temporal Analyses}
\label{sec:flow_and_temporal_analyses}

In addition to visualizing egocentric networks with a fixed radius, we can follow significant flows of value through the network over time. If a vertex representing a user receives a large volume of Bitcoins relative to their estimated balance, and, shortly after, transfers a significant proportion of those Bitcoins to another user, we deem this interesting. We built a special purpose tool that, starting with a chosen vertex or set of vertices, traces significant flows of Bitcoins over time. In practice we have found this tool to be quite revealing when analyzing the user network.

\emph{Case Study -- Part II:} To demonstrate this tool we re-consider the Bitcoin theft described earlier.
\ifthenelse {\boolean{longVersion}} {
We note that the victim has developed their own tool to generate an exhaustive list of public-keys that have received some portion of the stolen Bitcoins since the theft.\footnote{http://folk.uio.no/vegardno/allinvain-addresses.txt} However, this list grows very quickly and, at the time of writing, contained more than \numprint{34100} public-keys.
}

\begin{figure}[!ht]
\begin{center}
        \includegraphics[width=120mm]{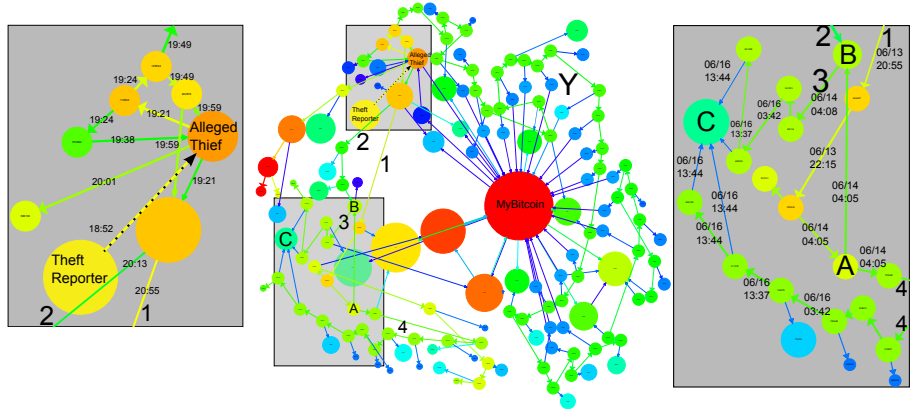}
\end{center}
\caption{Visualisation of Bitcoin flow from the alleged theft. The left inset shows the initial shuffling of Bitcoins among accounts close to that of the alleged thief. The right inset shows the flow of Bitcoins during several subsequent days. The flows split and ater merge, validating that the flows found by the tool are probably still controlled by a single user.}
\label{fig:visualisationTheft}
\end{figure}

Figure~\ref{fig:visualisationTheft} shows an annotated visualization produced using our tool. We observe several interesting flows in the aftermath of the theft. The initial theft of a small volume of 1 BTC is immediately followed by the theft of \numprint{25000} BTC. This is represented as a dotted black line between the relevant vertices, magnified in the left inset.

In the left inset, we can see that the Bitcoins are shuffled between a small number of accounts and then transferred back to the initial account. After this shuffling step, we have identified four significant outflows of Bitcoins that began at 19:49, 20:01, 20:13 and 20:55. Of particular interest are the outflows that began at 20:55 (labeled as `1' in both insets) and 20:13 (labeled as `2' in both insets). These outflows pass through several subsequent accounts over a period of several hours. Flow~1 splits at the vertex labeled $A$ in the right inset at 04:05 on the day after the theft. Some of its Bitcoins rejoin Flow~2 at the vertex labeled $B$. This new combined flow is labeled as `3' in the right inset. The remaining Bitcoins from Flow~1 pass through several additional vertices in the next two days. This flow is labeled as `4' in the right inset.

A surprising event occurs on 16/06/2011 at approximately 13:37. A small number of Bitcoins are transferred from Flow~3 to a heretofore unseen public-key $pk_1$.\footnote{1FKFiCYJSFqxT3zkZntHjfU47SvAzauZXN} Approximately seven minutes later, a small number of Bitcoins are transferred from Flow~3 to another heretofore unseen public-key $pk_2$.\footnote{1FhYawPhWDvkZCJVBrDfQoo2qC3EuKtb94} Finally, there are two simultaneous transfers from Flow~4 to two more heretofore unseen public-keys: $pk_3$\footnote{1MJZZmmSrQZ9NzeQt3hYP76oFC5dWAf2nD} and $pk_4$.\footnote{12dJo17jcR78Uk1Ak5wfgyXtciU62MzcEc} We have determined that these four public-keys, $pk_1$, $pk_2$, $pk_3$ and $pk_4$ -- which receive Bitcoins from two separate flows that split from each other two days previously -- are all contracted to the same user in our ancillary network. This user is represented as $C$ in Fig.~\ref{fig:visualisationTheft}.

There are several other examples of interesting flow. The flow labeled as $Y$ involves the movement of Bitcoins through thirty unique public-keys in a very short period of time. At each step, a small number of Bitcoins (typically 30 BTC which had a market value of approximately US\$500 at the time of the transactions) are siphoned off. The public-keys that receive the small number of Bitcoins are typically represented by small blue vertices due to their low volume and degree. On 20/06/2011 at 12:35, each of these public-keys makes a transfer to a public-key operated by the MyBitcoin service.\footnote{1MAazCWMydsQB5ynYXqSGQDjNQMN3HFmEu} Curiously, this public-key was previously involved in another separate Bitcoin theft\footnote{http://forum.bitcoin.org/index.php?topic=20427.0}. 

\ifthenelse {\boolean{longVersion}} {

\begin{figure}[!htb]
\begin{center}
        \includegraphics[width=60mm]{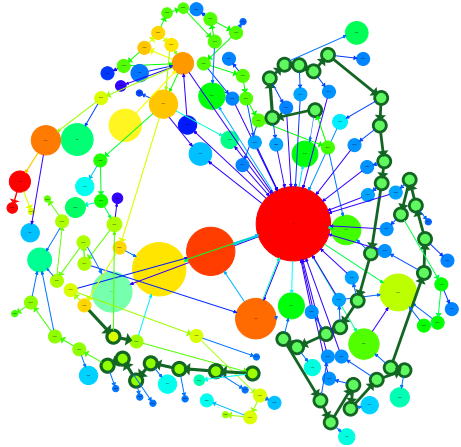}
\end{center}
\caption{The Bitcoins are transferred between public-keys along the highlighted paths very quickly.}
\label{fig:temporalLeak}
\end{figure}

We also observe that the Bitcoins in many of the above flows are transferred between public-keys very quickly. Fig.~\ref{fig:temporalLeak} shows two flows in particular where the intermediate parties waited for very few confirmations before re-sending the Bitcoins to other public-keys.

}

Much of this analysis is circumstantial. We cannot say for certain whether or not these flows imply a shared agency in both incidents. However, it does illustrate the power of our tool when tracing the flow of Bitcoins and generating hypotheses. It also suggests that a centralized service may have further details on the user(s) in control of the implicated public-keys.

\ifthenelse {\boolean{longVersion}} {

\subsection{Other Forms of Analysis}
\label{sec:other_forms_of_analysis}

There are many other forms of analysis that can be applied in order to de-anonymize the workings of the Bitcoin system:
\begin{itemize}
\item Many transactions have two outputs: one is the payment from a payer to a payee and the other is the return of change to the payer. If we assume that a transaction was created using a particular client implementation and we have access to the client's source code, then we may be able to deduce, in some cases, which was the output and which was the change. We can then map the public-key that the change was assigned to back to the user who created the transaction.
\item Order books for Bitcoin exchanges are typically available to support trading tools. As orders are often placed in Bitcoin values converted from other currencies, they have a precise decimal value with eight significant digits. It may be possible to find transactions with corresponding amounts and thus map public-keys and transactions to the exchanges.
\item Over an extended time period, several public-keys, if used at similar times, may belong to the same user. It may be possible to construct and cluster a co-occurrence network to help deduce mappings between public-keys and users.
\item Finally, there are far more sophisticated forms of attack where the attacker actively participates in the network, for example, using marked Bitcoins or by operating a laundry service.
\end{itemize}

}

\subsection{Mitigation Strategies}
\label{sec:mitigation_strategies}
In addition to educating users about the limits of anonymity in the Bitcoin system, some risks to privacy could potentially be mitigated by making changes to the system.
A patch to the official Bitcoin client has been developed\footnote{http://coderrr.wordpress.com/2011/06/30/patching-the-bitcoin-client-to-make-it-more-anonymous \dd\ Retrieved 2011-11-04} which allows users to prevent the linking of public-keys by making the user aware of potential links within the Bitcoin client user-interface. It is also possible for the client to automatically proxy Bitcoins through dummy public-keys.
This would come at the cost of increased transaction fees but would increase deniability and obfuscate the chain of transaction histories.
Finally, if a future version of the protocol supported protocol-level mixing of Bitcoins, this would increase the difficulty for a passive third-party to track individual user histories.

\section{Conclusions}
\label{sec:conclusions}

For the past half-century futurists have heralded the advent of a
cash-less society~\clubpenalty10000\cite{anderson-et-al-66}. Many of their
predictions have been realized, \eg
Anderson et al.'s~\clubpenalty10000\cite{anderson-et-al-66}'s `on-line real-time'
payment system and bank-maintained `central information files'.
However, cash is still a competitive and relatively anonymous means of
payment. Bitcoin is an electronic analog of cash in the online world. It is
decentralized: there is no central authority responsible for the issuance
of Bitcoins and there is no need to involve a trusted third-party when
making online transfers. However, this flexibility comes at a price: the
entire history of Bitcoin transactions is publicly available. In this chapter 
we investigated the structure of two networks derived from this dataset and
their implications for user anonymity.

Using an appropriate network representation, it is possible to associate many public-keys with each other, and with external identifying information. With appropriate tools, the activity of known users can be observed in detail. This can be performed using a passive analysis only. Active analyses, where an interested party can potentially deploy `marked' Bitcoins and collaborate with other users can discover even more information. We also believe that large centralized services such as the exchanges and wallet services are capable of identifying and tracking considerable portions of user activity.

Technical members of the Bitcoin community have cautioned that strong anonymity is not a prominent design goal of the Bitcoin system. However, casual users need to be aware of this, especially when sending Bitcoins to users and organizations they would prefer not to be publicly associated with.

\section{Acknowledgements}
This research was supported by Science Foundation Ireland (SFI) Grant No. 08/SRC/I1407: Clique: Graph and Network Analysis Cluster. Both authors contributed equally to this work. It was performed independently of any industrial partnership or collaboration of the Clique Cluster.

\bibliographystyle{abbrv}
\bibliography{main}

\end{document}